\documentclass[aps,prd,nofootinbib,reprint,superscriptaddress,preprintnumbers]{revtex4-2}
\pdfoutput=1
\usepackage[utf8]{inputenc}
\usepackage{amsmath,amssymb}
\usepackage{epsfig}
\usepackage{comment}
\usepackage{graphicx}
\usepackage{float}
\usepackage[colorlinks,citecolor=blue]{hyperref}
\usepackage{xcolor}
\usepackage{subfigure}
\usepackage{cleveref}
\usepackage{url}

\begin{document}

\preprint{CETUP-2023-022, FERMILAB-PUB-24-0317-T, IPPP/24/35}
\title{Pseudo-Dirac Neutrinos and Relic Neutrino Matter Effect \\
on the High-energy Neutrino Flavor Composition}
\author{P. S. Bhupal Dev}
\email{bdev@wustl.edu}
\affiliation{Department of Physics and McDonnell Center for the Space Sciences, Washington University, St. Louis, MO 63130, USA}
\author{Pedro A. N. Machado}
\email{pmachado@fnal.gov}
\affiliation{Theoretical Physics Department, Fermilab, P.O. Box 500, Batavia, IL 60510, USA}
\author{Ivan Mart\'\i{}nez-Soler}
\email{ivan.j.martinez-soler@durham.ac.uk}
\affiliation{Institute for Particle Physics Phenomenology, Durham University, South Road, DH1 3LE, Durham, UK}

\begin{abstract}
    We show that if neutrinos are pseudo-Dirac, they can potentially affect the flavor ratio predictions for the high-energy astrophysical neutrino flux observed by IceCube. In this context, we point out a novel matter effect induced by the cosmic neutrino background (C$\nu$B) on the flavor ratio composition. Specifically, the active-sterile neutrino oscillations over the astrophysical baseline lead to an energy-dependent flavor ratio at Earth due to the C$\nu$B matter effect, which is in principle distinguishable from the vacuum oscillation effect, provided there is an asymmetry between the neutrino and antineutrino number densities, as well as a local C$\nu$B overdensity. Considering the projected precision of the 3-neutrino oscillation parameter measurements and improved flavor triangle measurements, we show that the next-generation neutrino telescopes, such as KM3NeT and IceCube-Gen2, can in principle probe the pseudo-Dirac neutrino hypothesis and the C$\nu$B matter effect.   
\end{abstract}
\maketitle
\section{Introduction} \label{sec:intro}
Despite great progress in neutrino physics over the past decades, the nature of neutrino mass remains unknown. Neutrinos could be either Majorana or Dirac particles. Or they could be somewhere in-between, namely, pseudo-Dirac~\cite{Wolfenstein:1981kw,Petcov:1982ya,Valle:1983dk,Doi:1983wu,Kobayashi:2000md}, which are fundamentally Majorana fermions, but behave like Dirac particles in laboratory experiments because of the extremely small mass-squared splitting ($\delta m^2$) between the active and sterile components. 
The theoretical and model-building aspects of pseudo-Dirac neutrinos have been extensively discussed in the literature; see e.g., Refs.~\cite{Chang:1999pb, Nir:2000xn,Joshipura:2000ts,Lindner:2001hr, Balaji:2001fi,Stephenson:2004wv,McDonald:2004qx, deGouvea:2009fp, Ahn:2016hhq, Joshipura:2013yba, Babu:2022ikf, Carloni:2022cqz}. 
In fact, in any model where the neutrinos start as
Dirac particles with naturally small masses 
could actually receive quantum gravity corrections making them pseudo-Dirac particles at a more fundamental level.
These corrections will generate small $\delta m^2$  via higher-dimensional operators suppressed by the Planck scale.
It is interesting to note that certain string landscape (swampland) constructions also predict pseudo-Dirac neutrinos~\cite{Ooguri:2016pdq, Ibanez:2017kvh, Gonzalo:2021zsp, Casas:2024clw}. 
Small $\delta m^2$ values could also be linked to the observed baryon asymmetry of the Universe~\cite{Ahn:2016hhq, Fong:2020smz}. Recently, the pseudo-Dirac neutrinos were also shown to resolve the excess radio background issue~\cite{Chianese:2018luo, Dev:2023wel}.

Irrespective of the theoretical motivations, the only experimental way to directly probe the active-sterile oscillations of pseudo-Dirac neutrinos with tiny mass splittings is by going to extremely long baselines, which is possible with astrophysical sources of neutrinos, such as solar~\cite{Giunti:1992hk, Cirelli:2004cz, deGouvea:2009fp,Anamiati:2017rxw, deGouvea:2021ymm, 
Ansarifard:2022kvy, Franklin:2023diy}, supernova~\cite{DeGouvea:2020ang, Martinez-Soler:2021unz}, high-energy astrophysical~\cite{2000ApJS,2002ApJS, Beacom:2003eu, Keranen:2003xd, Esmaili:2009fk,Esmaili:2012ac,Joshipura:2013yba, Shoemaker:2015qul, Brdar:2018tce, Carloni:2022cqz}, or relic neutrinos~\cite{Perez-Gonzalez:2023llw}. 
In fact, stringent upper limits on $\delta m^2_{1,2}\lesssim 10^{-12}~{\rm eV}^2$ have been derived using the solar neutrino data~\cite{deGouvea:2009fp, Ansarifard:2022kvy}. There also exists an old limit on $\delta m^2_i\lesssim 10^{-8}~{\rm eV}^2$ from Big Bang Nucleosynthesis (BBN) considerations~\cite{Barbieri:1989ti, Enqvist:1990ek}.
These limits are derived assuming the usual maximal active-sterile neutrino mixing in the pseudo-Dirac scenario. 
If the mixing is non-maximal, the $\delta m^2$ limits can be much  weaker~\cite{Chen:2022zts}. 
Moreover, the solar neutrino data is not sensitive to $\delta m^2_3$ due to the smallness of $\theta_{13}$, and the limit from atmospheric, accelerator and reactor neutrino data is rather weak, $\delta m^2_3\lesssim 10^{-5}~{\rm eV}^2$~\cite{Anamiati:2017rxw}, due to the much shorter baselines.

The recent identification of a few point sources for astrophysical neutrinos~\cite{IceCube:2022der} allowed us to set the first IceCube limits on the pseudo-Dirac neutrino hypothesis in the $\delta m^2_i\in [10^{-21},10^{-16}]~{\rm eV}^2$ range~\cite{Carloni:2022cqz}; see also Refs.~\cite{Rink:2022nvw, Dixit:2024ldv} for related analyses. 
However, these studies only used the IceCube track-like sample (mostly involving muon neutrinos, with a small fraction coming from tau-induced tracks), and hence, were insensitive to the full neutrino flavor information. 
This is justifiable because the track events have excellent angular resolution of $\lesssim  0.2^{\circ}$~\cite{IceCube:2013uto} and are therefore ideal for point source identification~\cite{IceCube:2021xar}, unlike the cascade events which have a poor angular resolution of $\sim 10^\circ$--$15^\circ$ at IceCube~\cite{IceCube:2017der}. 
The cascade resolution will significantly improve up to $1.5^\circ$ at KM3NeT~\cite{KM3Net:2016zxf} with their current high-energy cascade reconstruction algorithm, and even sub-degree resolution can be achieved with better reconstruction algorithms using the timing information and elongation emission profile of cascades~\cite{vanEeden:2021zzv}.

In this paper, we study how including the cascade events can give us additional information on the pseudo-Dirac neutrino hypothesis. 
In particular, we show that the flavor ratio measurements of high-energy neutrinos, from either diffuse or point sources, would be affected in the presence of pseudo-Dirac neutrinos, except for the special case when all three active-sterile mass splittings are exactly the same. 
Given the fact that the flavor ratio measurements are expected to improve significantly~\cite{Song:2020nfh} with KM3NeT~\cite{KM3Net:2016zxf} and Baikal-GVD~\cite{Allakhverdyan:2021vkk} currently under construction, and with the proposed next-generation neutrino telescopes, such as IceCube-Gen 2~\cite{IceCube-Gen2:2020qha},  P-ONE~\cite{P-ONE:2020ljt}, NEON~\cite{Zhang:2024slv}, HUNT~\cite{Huang:2023mzt}, TRIDENT~\cite{Ye:2022vbk}, TAMBO~\cite{Thompson:2023pnl}, Trinity~\cite{Otte:2019aaf} and RET~\cite{RadarEchoTelescope:2021czz}, they will provide an unprecedented opportunity to test the  pseudo-Dirac neutrino hypothesis.

The final flavor ratio measured on Earth crucially depends on the initial source flavor composition which is currently unknown. 
We take this into account by considering different well-motivated choices for $(\nu_e:\nu_\mu:\nu_\tau)$ at the source,\footnote{Since IceCube cannot  distinguish between neutrinos and antineutrinos on an event-by-event basis (with the exception of the Glashow resonance~\cite{Glashow:1960zz} for which we lack statistics~\cite{IceCube:2021rpz}), we take the sum of neutrinos and antineutrinos for a given flavor.} namely,  (i) $(1/3:2/3:0)$ for the standard pion and muon decay~\cite{Learned:1994wg}; 
(ii) $(0:1:0)$ for the muon damped case~\cite{Rachen:1998fd, Kashti:2005qa, Kachelriess:2007tr, Hummer:2010ai, Winter:2014pya}; 
(iii) $(1:0:0)$ for neutron decay~\cite{Anchordoqui:2003vc, Anchordoqui:2014pca}; and 
(iv) $(x:1-x:0)$ with $x\in [0,1]$ for the general case corresponding to a mixture of multiple processes/sources contributing to the neutrino flux. 
In each case, we compare the expectations from the standard 3-neutrino oscillation paradigm with the pseudo-Dirac scenario for a given $\delta m^2$ to see whether they can be distinguished from each other on the flavor triangle. 
Note that since we are dealing with flavor ratios, we are insensitive to uncertainties related to the normalization or energy dependence of the astrophysical neutrino flux.

Moreover, for the $\delta m^2$ values of interest here, we show that the matter effect due to the cosmic neutrino background (C$\nu$B) can play an important role in determining the flavor ratios on Earth, depending on the value of the neutrino-antineutrino asymmetry and the local C$\nu$B overdensity. 
This is in contrast with the pure vacuum oscillations assumed so far in the vast literature of flavor ratio studies (see e.g., Refs.~\cite{Lipari:2007su, Mena:2014sja, Palladino:2015zua, Bustamante:2015waa, Shoemaker:2015qul, Brdar:2018tce, Bustamante:2019sdb, Palladino:2019pid, Song:2020nfh, Liu:2023flr}).\footnote{The effect of source matter effect on the flavor composition of high-energy neutrinos from active galactic nuclei was recently considered in Ref.~\cite{Dev:2023znd}, but this becomes important only for heavily Compton-thick sources with column density $\gtrsim 10^{30}~{\rm cm}^{-2}$ whose exact population or contribution to the observed flux at IceCube is currently unknown.} 
Only the left-handed component of the pseudo-Dirac neutrino actively interacts with the Standard Model (SM) fermions via  weak interactions, whereas the right-handed component is sterile. Thus, the neutral-current interactions of the left-handed component of the high-energy neutrino flux with the C$\nu$B bath would induce a difference in the matter potential for a given flavor (depending on which $\delta m_i^2\neq 0$), which could modify the oscillation probabilities, and could even induce an MSW resonance~\cite{Wolfenstein:1977ue, Mikheyev:1985zog} for suitable values of $\delta m_i^2$. 
Note that in the standard 3-neutrino case, the neutral-current interaction equally affects all three flavors and does not lead to a matter potential difference between different flavors, unless there is an asymmetry between neutrinos and antineutrinos.
The same is true if all three active-sterile mass splittings $\delta m^2_i$ are the same, in which case there is no matter potential difference induced by C$\nu$B either. 
Thus, including the C$\nu$B matter effect would provide an additional handle on probing small $\delta m_i^2$ values at neutrino telescopes. 
Moreover, the matter effect introduces a novel energy-dependent flavor transition, which will help us disentangle the pseudo-Dirac scenario. 

The rest of the paper is organized as follows: In Section~\ref{sec:standard}, we review the standard 3-flavor oscillation paradigm for the flavor triangle analysis. In Section~\ref{sec:vac}, we present the pseudo-Dirac case with oscillations in vacuum and in Section~\ref{sec:matter} the oscillations in matter, both cases in a time-independent background. In Section~\ref{sec:Hubble}, we discuss the oscillations both with and without C$\nu$B matter effect in an expanding Universe. In Section~\ref{sec:over}, we include the C$\nu$B overdensity and the finite cluster size effect. Our results are given in Section~\ref{sec:results}. We conclude with some final remarks in Section~\ref{sec:con}.

\section{Standard Case}
\label{sec:standard}
In the standard 3-neutrino oscillation scenario, the neutrino flavor eigenstates $|\nu_\alpha\rangle$, with $\alpha=e,\mu,\tau$, are  related to the mass eigenstates $|\nu_i\rangle$, with $i=1,2,3$, via a unitary transformation, i.e. 
\begin{align}
|\nu_\alpha\rangle =\sum_{i=1}^3 U_{\alpha i}^*|\nu_i\rangle \, ,
\end{align}
where $U$ is the $3\times 3$ Pontecorvo-Maki-Nakagawa-Sakata (PMNS) lepton mixing matrix, parameterized in terms of three mixing angles $\theta_{ij}$ and a Dirac CP phase $\delta_{\rm CP}$~\cite{ParticleDataGroup:2022pth}.\footnote{If neutrinos are Majorana, $U$ contains two additional phases which however do not affect oscillations.} 

The characteristic neutrino oscillation length scale in vacuum is given by  
\begin{align}
\label{eq:Losc1} 
    L^{\rm std}_{\rm osc}& =\frac{4\pi E_\nu}{\Delta m^2_{ij}} 
    \simeq 8\times 10^{-6}{\rm pc} \left(\frac{E_\nu}{1~{\rm TeV}}\right)\left(\frac{10^{-5}~{\rm eV}^2}{\Delta m_{ij}^2}\right),  
\end{align}
where $E_\nu\gg m_i$ is the neutrino energy and $\Delta m_{ij}^2\equiv |m_i^2-m_j^2|$ are the mass-squared differences. From this equation, 
it is clear that for high-energy neutrinos, $L_{\rm osc}$ corresponding to either solar or atmospheric mass-squared splitting 
is much smaller than the typical distance ($\gtrsim$ Mpc) to the extragalactic astrophysical sources. 
Therefore, the standard 3-neutrino oscillations are rapid enough to average out over astrophysical baselines, and we are only sensitive to the averaged out $\nu_\alpha\to \nu_\beta$ flavor transition probability,  
\begin{align}
    P_{\alpha\beta}^{\rm std} = \sum_{i=1}^3|U_{\alpha i}|^2|U_{\beta i}|^2 \, ,
    \label{eq:probstd}
\end{align}
which depends on the 3-neutrino mixing angles, as well as on the Dirac CP phase to a lesser extent. 
When drawing the allowed regions in the flavor triangles, we will use the best-fit and 68\% confidence level (CL) allowed values of the oscillation parameters from the recent {\tt NuFit 5.3} global fit~\cite{Esteban:2020cvm, nufit}, assuming a normal mass ordering for concreteness. Note that the latest oscillation results from T2K~\cite{t2k} and NO$\nu$A~\cite{nova} individually continue to show a mild preference for normal mass ordering, although their combination prefers inverted mass ordering~\cite{Kelly:2020fkv}, so this is still an open question. 

Thus, for a given initial flavor composition at the source $(f_{e},f_\mu,f_\tau)_{\rm S}$, the final flavor composition at Earth  under standard vacuum oscillations is given by 
\begin{align}
    f_{\beta,\oplus} = \sum_{\alpha=e,\mu,\tau} P_{\alpha\beta}^{\rm std} f_{\alpha,{\rm S}} \, ,
    \label{eq:fstd}
\end{align}
where we have normalized the flavor ratios so that they add up to unity, i.e., $\sum _{\alpha}f_{\alpha,{\rm S}}=\sum _{\beta}f_{\beta,\oplus}=1$. Depending on the physical scenario for the initial source flavor composition, we can then calculate the final flavor composition at Earth using Eq.~\eqref{eq:fstd}. This will be referred to as the ``standard case" in the following.

\section{Pseudo-Dirac Case}
\label{sec:vac}

A simple extension of the SM that explains the nonzero neutrino masses consists of the addition of $n$ SM-singlet Weyl fermions $N^i$, usually referred to as right-handed (or sterile) neutrinos. It gives rise to two additional terms in the Lagrangian:
\begin{equation}
    -\mathcal{L} \supset \lambda_{\alpha i}L^{\alpha} H N^{i} + \sum_{i,j=1}^n \frac{M_{ij}}{2}N^{i}N^{j} + {\rm H.c.} \, , 
\label{eq:lag}
\end{equation}
where $L$ and $H$ are the $SU(2)_L$-doublet lepton and Higgs fields respectively, $\lambda_{\alpha i}$ are the Dirac neutrino Yukawa couplings and $M_{ij}$ are the Majorana masses for the right-handed neutrinos, which break the global lepton number symmetry of the SM. Note that $M$ is a symmetric matrix, i.e., $M_{ij}=M_{ji}$ by construction, and we need $n\geq 2$ to reproduce the observed neutrino oscillation data with two nonzero mass-squared splittings. For concreteness, we choose $n=3$, i.e., 3 right-handed neutrinos. 

After electroweak symmetry breaking by the vacuum expectation value of the neutral component of the Higgs field, $\langle H^0\rangle=v$, Eq.~\eqref{eq:lag} describes $3+3$ Majorana neutrinos, given by the $6\times 6$ mass matrix
\begin{equation}
     {\cal M} = \left(\begin{array}{cc}
 {\bf 0} &m\\
 m^T &M\\
 \end{array}\right)  \, , 
 \label{eq:mass}
\end{equation}
where $m=\lambda v$. We can choose the weak basis to diagonalize the Dirac mass term by a unitary matrix $U$, i.e. $m=U^*m_{\rm diag}$. The symmetric matrix $M$ is diagonalized by another unitary matrix $U_R$, i.e. $M=M^T=U_R^TM_{\rm diag}U_R$. When $M={\bf 0}$, the 6 mass eigenstates in Eq.~\eqref{eq:mass} form 3 Dirac pairs with squared masses equal to the eigenvalues of $m_D^\dag m_D$. Therefore, the limit $M\ll m$ (in the eigenvalue sense) refers to the pseudo-Dirac scenario, where the mass splitting induced by the Majorana mass term $M$ is small compared to the leading-order Dirac mass term $m$ in the eigenvalues of the mass matrix~\eqref{eq:mass}.  

In the pseudo-Dirac limit, to leading-order, Eq.~\eqref{eq:mass} can be written as~\cite{deGouvea:2009fp}
\begin{align}
    {\cal M} \simeq & \begin{pmatrix} {\bf 1} & -\delta^* \\
    \delta^T & {\bf 1} \end{pmatrix}   \frac{1}{\sqrt 2}\begin{pmatrix}
        U^* & -U^* \\ {\bf 1} & {\bf 1} 
    \end{pmatrix}\nonumber \\
    & \times 
    \begin{pmatrix}
        m_{\rm diag}({\bf 1}+\epsilon_{\rm diag}) & {\bf 0} \\
        {\bf 0} & -m_{\rm diag}({\bf 1}-\epsilon_{\rm diag})
    \end{pmatrix}\nonumber \\
    & \times 
    \frac{1}{\sqrt 2}\begin{pmatrix}
        U^\dag & {\bf 1} \\ -U^\dag & {\bf 1} 
    \end{pmatrix}
    \begin{pmatrix} {\bf 1} & \delta \\
    -\delta^\dag & {\bf 1} \end{pmatrix} \, ,
\end{align}
where $\delta$ is a dimensionless matrix of small numbers:
\begin{align}
    \delta = U\left(\frac{\epsilon_{\rm diag}}{2}+\varepsilon\right) \, , 
\end{align}
with the diagonal entries of $\varepsilon$ vanishing, and $(\epsilon_{\rm diag},\varepsilon)$ are functions of $(M,m_{\rm diag})$: 
\begin{align}
    M & = 2\epsilon_{\rm diag}m_{\rm diag}+\varepsilon^Tm_{\rm diag}+m_{\rm diag}\varepsilon \, ,  \label{eq:eps1} \\
    {\rm and}~\quad & m_{\rm diag}\varepsilon^T = -\varepsilon m_{\rm diag} \, .
    \label{eq:eps2}
\end{align}
From Eq.~\eqref{eq:eps2}, it is clear that $m_{\rm diag}$ and $\varepsilon$ do not commute. Eq.~\eqref{eq:eps1} implies that $M$ is indeed a symmetric matrix, whose diagonal entries proportional to $\epsilon_{\rm diag}$ determine the mass-squared splitting between the quasi-degenerate states, while its off-diagonal entries proportional to $\varepsilon$ contribute only to the $6\times 6$ mixing matrix that diagonalizes ${\cal M}_\nu^\dag {\cal M}_\nu$:  
\begin{align}
    \widetilde{U} & = \frac{1}{\sqrt 2}\begin{pmatrix}
        {\bf 1} & -\delta \\
        \delta^\dag & {\bf 1}
    \end{pmatrix}
    \begin{pmatrix}
        U & -U \\
        {\bf 1} & {\bf 1}
    \end{pmatrix}\nonumber \\
    & = \frac{1}{\sqrt 2}\begin{pmatrix}
        U\left({\bf 1}-\frac{\epsilon_{\rm diag}}{2}-\varepsilon\right) & - U\left({\bf 1}+\frac{\epsilon_{\rm diag}}{2}+\varepsilon\right) \\
        {\bf 1}+\frac{\epsilon_{\rm diag}}{2}-\varepsilon & {\bf 1}-\frac{\epsilon_{\rm diag}}{2}+\varepsilon
    \end{pmatrix}\, .
\end{align}
Here we have defined the flavor eigenstates as $\nu_\alpha$ (with $\alpha=e,\mu,\tau,s_1,s_2,s_3)$ and the mass eigenstates as $\nu_i$ (with $i=1a, 2a, 3a, 1s, 2s, 3s)$, with $a$ ($s$) denoting the mostly active (sterile) states. The states are ordered in the usual way, i.e. for normal ordering, we have $m_{1a}^2<m_{2a}^2<m_{3a}^2$ and similarly for the sterile states. In the pseudo-Dirac limit, we assume that the $\epsilon_{\rm diag}$ and $\varepsilon$ parameters are small enough so that the mass-squared splitting $\delta m_i^2\equiv |m_{ia}^2-m_{is}^2|\ll m_i^2$ for all $i=1,2,3$.

The Hamiltonian describing the neutrino evolution in vacuum is thus given by $H^{\rm PD}_{\text{vac}} = \widetilde{U}^{\dagger} {\cal M}_{\text{diag}}^2 \widetilde{U}/2E_\nu$, where the masses can be separated into two sub-block $3\times 3$ diagonal matrices $M_{\text{diag}}^2=\{m^2_{iS},m^2_{iA}\}$, with the squared mass eigenvalues  
\begin{align}
m^2_{iS} & = m^2_{i} + \delta m^2_{i}/2\, , \\
m^2_{iA} & = m^2_{i} - \delta m^2_{i}/2 \, ,
\end{align}
corresponding to the mass eigenstates 
\begin{align}
    \nu_{iS} & = \sin\theta_i \nu_{ia}+\cos\theta_i \nu_{is} \, , \label{eq:4} \\
    \nu_{iA} & = -i(\cos\theta_i \nu_{ia}-\sin\theta_i \nu_{is}) \, , \label{eq:5}
\end{align}
with $\nu_{ia}$ and $\nu_{is}$ being the active and sterile components, respectively. In the case of pseudo-Dirac states, a maximal mixing between $\nu_{S}$ and $\nu_{A}$ states is assumed, i.e., $\theta_i=\pi/4$, in which case the states coincide with the symmetric ($\nu_S=(\nu_a+\nu_s)/\sqrt 2$) and anti-symmetric ($\nu_A=-i(\nu_a-\nu_s)/\sqrt 2$) combinations of the active and sterile components. Therefore, the mixing matrix is simply given by
\begin{equation}
 \widetilde{U} = \frac{1}{\sqrt{2}}\left(\begin{array}{cc}
 U &0_{3\times3}\\
 0_{3\times3} &U_{R}\\
 \end{array}\right) \left(\begin{array}{cc}
 1_{3\times 3} &i_{3\times3}\\
 1_{3\times3} &-i_{3\times3}\\
 \end{array}\right)   \, , 
 \label{eq:16a}
\end{equation}
where $U$ is identified as the PMNS matrix and $U_{R}$ is the mixing matrix between the right-handed (sterile) states. In this maximal mixing limit, the off-diagonal part of $M$ in Eq.~\eqref{eq:eps1} which does not commute with $m$ can be safely ignored.  $\widetilde{U}$ in Eq.~\eqref{eq:16a} also depends on three additional phases, one for each pair of the pseudo-Dirac states~\cite{Anamiati:2017rxw, Anamiati:2019maf}, which are assumed to be zero in this case.

\section{C$\nu$B Matter Effect}
\label{sec:matter}
The interactions of high-energy neutrinos of a given flavor $\alpha$ with the C$\nu$B of a given flavor $\beta$ introduce a matter potential, given by the $6\times 6$ matrix $\widehat{V}_{\nu_\alpha} = V_{\nu_\alpha} \text{diag}\{{\bf 1}_{3\times 3}, {\bf 0}_{3\times 3}\}$,\footnote{For simplicity, we assume that the C$\nu$B matter effect is flavor-universal. This is certainly valid if the C$\nu$B contains neutrinos of all flavor with equal number densities and if they interact only via weak interactions. 
Decaying neutrinos~\cite{Valera:2024buc, Batell:2024hzo} or the presence of flavor-nonuniversal nonstandard interactions~\cite{Berryman:2022hds} would need special treatment.} where~\cite{Notzold:1987ik}
\begin{align}
    V_{\nu_\alpha} = \sqrt 2 G_F (1+\delta_{\alpha\beta})(n_{\nu_\beta}-n_{\bar{\nu}_\beta}) 
    \label{eq:pot}
\end{align}
at leading order, and $n_{\nu_\beta(\bar{\nu_\beta})}$ denote the background neutrino (antineutrino) number densities, and  $G_F$ is the Fermi constant. The $\delta_{\alpha\beta}$ term arises from the exchange between neutrinos of the same flavor.  For high-energy antineutrinos, Eq.~\eqref{eq:pot} has an overall sign difference. It is clear that the matter effect vanishes at leading order unless there is an asymmetry between the background neutrino and antineutrino number densities. This is reminiscent of the Stodolsky effect~\cite{Stodolsky:1974aq}, namely, a spin-dependent energy shift induced by the C$\nu$B background, which also crucially depends on a nonzero neutrino asymmetry. Note that there is another term in the matter potential which is proportional to $(n_{\nu_\beta}+n_{\bar{\nu}_\beta})$~\cite{Notzold:1987ik}, but it comes with a $G_F^2$ suppression, and therefore, is negligible compared to Eq.~\eqref{eq:pot} with an ${\cal O}(1\%)$ asymmetry.

A primordial neutrino asymmetry existing prior to neutrino decoupling epoch leaves imprints on the cosmic evolution, which is constrained by various cosmological observables, including BBN, Cosmic Microwave Background (CMB) and large-scale structure (LSS). The cosmological constraints on the asymmetry can be parameterized in terms of the degeneracy parameters $\xi_\alpha\equiv \mu_\alpha/T$ (where $\mu_\alpha$'s are the chemical potentials) by~\cite{Serpico:2005bc} 
\begin{align}
    \eta_{\nu_\alpha} \equiv  & \frac{n_{\nu_\alpha}-n_{\bar\nu_\alpha}}{n_\gamma} \nonumber \\
    = & \frac{1}{12\zeta(3)}\left(\frac{T_{\nu_\alpha}}{T_\gamma}\right)^3 \pi^2\xi_{\nu_\alpha}\left(1+\frac{\xi^2_{\nu_\alpha}}{\pi^2}\right) \simeq 0.25 \xi_{\nu_\alpha}\, . 
    \label{eq:eta}
\end{align}
In fact, the recent near-infrared observations of extremely metal-poor galaxies in the EMPRESS survey of the Subaru imaging data gives a precise determination of the primordial $^4$He abundance, which is in $>2\sigma$ tension with previous $N_{\rm eff}$ results and prefers a non-zero neutrino chemical potential~\cite{Matsumoto:2022tlr}. Subsequent joint analyses of the BBN, CMB and LSS data also found $\sim 2\sigma$ evidence for a neutrino asymmetry~\cite{Burns:2022hkq, Escudero:2022okz, Li:2024gzf}, with the current best-fit value of  $\xi_{\nu_e}=0.024\pm 0.013$~\cite{Li:2024gzf}. If one does not consider the EMPRESS result, the reduced chemical potential at BBN is constrained to be $\eta_{\nu}\lesssim 10^{-2}$~\cite{Serpico:2005bc, Simha:2008mt}. However, this is strictly true only for the chemical potential of electron-flavor (anti)neutrinos. It has been held that neutrino flavor oscillations, active below $\sim$ 10 MeV, would bring initially different asymmetries to the same value, hence making the $10^{-2}$  constraint true for all flavors.
This ``perfect equilibration" is, however, not true in general and is an artifact of using a damping approximation for the collision term~\cite{Froustey:2021azz}. Initially different asymmetries get mixed, but not perfectly. Importantly, at BBN, the electron neutrino chemical potential is constrained to be very small (about 0.03--0.04 depending on the BBN code used when including the EMPRESS result), but muon and tau chemical potentials can be much larger. As shown in an example situation in Ref.~\cite{Froustey:2024mgf}, the final electron asymmetry is very small, while the mu/tau asymmetry is of order 0.1. The only constraint on too large asymmetries in the muon/tau sector actually comes from $N_{\rm eff}$, but it is still poorly constrained by current CMB experiments. 

From a theoretical viewpoint, one might naively expect that due to the sphaleron processes in the early universe~\cite{Kuzmin:1985mm, Harvey:1990qw}, the neutrino asymmetries, or lepton asymmetries in general, should be of the same order as the baryon asymmetry, $\eta_B\equiv (n_B-n_{\bar B})/n_\gamma=(6.14\pm 0.04)\times 10^{-10}$~\cite{Planck:2018vyg}.  However, there exist various models in which a substantially large lepton asymmetry compared to the baryon asymmetry can still be generated before the neutrino decoupling epoch; see e.g., Refs.~\cite{Affleck:1984fy, Dolgov:1989us, Foot:1995qk, Casas:1997gx, Bajc:1997ky, Akhmedov:1998qx, McDonald:1999in,  March-Russell:1999hpw, Kawasaki:2022hvx, Domcke:2022uue,  Borah:2022uos, ChoeJo:2023cnx, Borah:2024xoa}.\footnote{A large lepton asymmetry is also required in the Shi-Fuller mechanism for resonant production of sterile neutrino dark matter~\cite{Shi:1998km}.} Without resorting to any specific model, we will simply assume an ${\cal O}(1\%)$ asymmetry in Eq.~\eqref{eq:pot}. Additionally, we assume that the asymmetry is the same for all flavors, which means that the $\delta_{\alpha\beta}$ appearing in Eq.~\eqref{eq:pot} is always 1. Otherwise, we would have a potential that depends on the neutrino flavor leading to additional flavor effects that are beyond the scope of this work.

To diagonalize the new Hamiltonian $H^{\rm PD}_{\rm mat}=H^{\rm PD}_{\rm vac}+\widehat{V}_\nu$ in the presence of the matter potential for the pseudo-Dirac case, we notice that $\widehat{V}_\nu$ commutes with both $U$ and $U_{R}$. 
Therefore, we can use three rotation matrices, one for each pair of degenerate states. 
The effective mixing angle in matter is given by
\begin{equation}
    \tan 2\widetilde{\theta}_{i} =  \frac{\delta m_i^2 \sin{(2\theta_i)}}{\delta m_i^2\cos{(2\theta_i)}-A} \, , 
    \label{eq:9}
\end{equation}
where $A=2E_\nu V_\nu$. Note that for non-maximal mixing, the standard Mikheyev-Smirnov-Wolfenstein (MSW) resonance condition~\cite{Wolfenstein:1977ue, Mikheyev:1985zog} would have been obtained when $A=\delta m_i^2\cos(2\theta)$. But in the pseudo-Dirac case with maximal mixing to start with, the matter effect tends to take the effective mixing angle away from the maximal value of $\pi/4$, as shown in Eq.~\eqref{eq:9}.

According to the $\Lambda$CDM model of cosmology, the C$\nu$B number density today is given by
\begin{align}
   n_{\nu,0}=\frac{3}{4}\frac{\zeta(3)}{\pi^2}g_\nu T^3_{\nu,0}\simeq 112~{\rm cm}^{-3} 
   \label{eq:11}
\end{align}
per neutrino flavor and the same for antineutrinos. Here $T_{\nu,0}=(4/11)^{1/3}T_{\gamma,0}\simeq 1.7\times 10^{-4}$ eV is the C$\nu$B temperature and $g_\nu=2$ is the number of degrees of freedom for each pseudo-Dirac neutrino. This gives a tiny matter potential $V_\nu\simeq 1.1\times 10^{-36}$ eV [cf.~Eq.~\eqref{eq:pot}] which, however, becomes relevant for $\delta m^2\gtrsim 2E_\nu V_\nu \simeq 2.2\times 10^{-24}~{\rm eV}\: (E_\nu/1~{\rm TeV})$.

In the presence of C$\nu$B matter effect, the  eigenvalues ($\lambda_{i S}, \lambda_{i A}$) of the diagonal matrix $M^2_{\rm diag}$ are given by
\begin{align}
    \lambda_{i S} & = \frac{A}{2}\cos 2\widetilde{\theta}_{i} + m^2_{i} + \frac{\delta m^2_{i}}{2} \sin 2\widetilde{\theta}_{i}\, , \\
    \lambda_{i A} & = -\frac{A}{2}\cos 2\widetilde{\theta}_{i} + m^2_{i} - \frac{\delta m^2_{i}}{2}\sin 2\widetilde{\theta}_{i} \, .
\end{align}
In the limit when the matter potential is negligible, i.e. $A \ll \delta m_i^2$, we recover maximal mixing between active and sterile neutrinos: $\widetilde{\theta}_{i} \to \pi/4$ [cf.~Eq.~\eqref{eq:9}] and the usual eigenvalues $\lambda_{i S} = m^2_{i1} + \delta m^2_{i}/2$ and $\lambda_{i A} = m^2_{i1} - \delta m^2_{i}/2$ [cf.~Eqs.~\eqref{eq:4} and \eqref{eq:5}]. For very large matter potentials, on the other hand, the mixing between $\nu_{i S}$ and $\nu_{i A}$ decreases, reaching the limit $\lambda_{i S} = A/2 +  m^2_{i}/2$ and $\lambda_{i A} = -A/2 +  m^2_{i}/2$.

In the scenario where the matter potential is constant along the neutrino evolution path, we can find the neutrino oscillation probability using the mixing angles and the eigenvalues from above. Considering the $\nu_\alpha\to \nu_\beta$ oscillation probability between the active states, we get
\begin{align}
    P_{\alpha\beta} & = \sum_j \left|U_{\alpha j}U^{\dagger}_{\beta j} \left[\cos^2\widetilde{\theta}_{j}\exp\left(\frac{-i\lambda_{jS}L}{2E_\nu}\right) \right.\right. \nonumber \\
    & \qquad \left. \left. + \sin^2\widetilde{\theta}_{j}\exp\left(\frac{-i\lambda_{jA}L}{2E_\nu}\right) \right]\right|^2 \, ,
\end{align}
where $L$ is the propagation length. 
The oscillation length induced by the active-active mass splitting $\Delta m^2_{j1}$, which is equal to $\Delta m^2_{\rm sol}\simeq 7.4\times 10^{-5}~{\rm eV}^2$ for $j=2$ and $\Delta m^2_{\rm atm}\simeq 2.5\times 10^{-3}~{\rm eV}^2$ for $j=3$~\cite{nufit}, is much shorter than the distance traveled by astrophysical neutrinos [cf.~Eq.~\eqref{eq:Losc1}] and is impossible to be resolved by the present detectors. 
Therefore, we average over it, thus obtaining
\begin{align}
   P_{\alpha\beta} & = \sum_{j} |U_{\alpha j}|^2|U_{\beta j}|^2\left[\cos^4\widetilde{\theta}_{j} + \sin^4\widetilde{\theta}_{j}  \right. \nonumber \\
   & \left. \qquad +2 \cos^2\widetilde{\theta}_{j}\sin^2\widetilde{\theta}_{j}\cos\left( \frac{\delta\widetilde{m}^2_{j} L}{4E_{\nu}} \right) \right] \, ,
   \label{eq:15}
\end{align}
where the effective mass-squared splitting in the presence of matter effect is given by 
\begin{align}
    \delta\widetilde{m}_j^2 & = \sqrt{A^2-2A\delta m_j^2\cos{(2\theta_j)}+(\delta m_j^2)^2} \nonumber \\
    & \simeq \sqrt{(\delta m_j^2)^2 + A^2} \, ,
    \label{eq:16}
\end{align}
which reduces to the vacuum mass-squared splitting $\delta m_i^2$ when $A\ll \delta m_j^2$, as expected. The effective active-sterile oscillation length scale in the presence of matter is 
\begin{align}
    L_{\rm osc} = \frac{4\pi E_\nu}{\delta \widetilde{m}_j^2} \, ,
    \label{eq:Losc}
\end{align} 
which now explicitly depends on the matter potential via Eq.~\eqref{eq:16}. It reduces to the vacuum case [cf.~Eq.~\eqref{eq:Losc1} with $\Delta m^2\to \delta m^2$] when $A\ll \delta m_i^2$. 

One might wonder whether the interactions of the high-energy neutrinos with the free electrons in the intergalactic medium (IGM) could also induce additional matter effect for the small $\delta m^2$ values under consideration. The mean IGM electronic density is $n_e\sim 10^{-7}~{\rm cm}^{-3}$~\cite{Dalton:2021afc}, which corresponds to a matter potential $V_e=\sqrt 2 G_F n_e \sim 10^{-44}$ eV. Even for a PeV-energy neutrino (the highest energy observed by IceCube), such a tiny matter potential will only be relevant if $\delta m^2\sim 10^{-29}~{\rm eV}^2$. However, in this case, the corresponding effective oscillation length is way beyond the size of the observable Universe, as we will see later. Therefore, we can safely neglect the IGM matter effect and only consider the C$\nu$B matter effect.

\section{Pseudo-Dirac neutrinos in expanding universe}
\label{sec:Hubble}

As the universe expands, the neutrino density from the C$\nu$B reduces. Considering that the neutrino density scales with the redshift as $n_\nu=n_{\nu, 0}(1+z)^3$, we have a matter potential that changes with redshift,  or in other words, with time. 

To estimate whether the neutrino evolution in an expanding universe is adiabatic or not, we have to compare the inverse oscillation length ($\delta \widetilde{m}^2/2E_\nu$)  with the transition between the massive states that is proportional to the variation of the effective mixing angle in matter ($d\widetilde{\theta} / dx $). Defining the adiabaticity parameter ($\gamma$) as the ratio between these two quantities~\cite{Mikheyev:1985zog}, we have 
\begin{equation}
    \gamma = \frac{\delta\widetilde{m}^2}{2E_{\nu}}\frac{1}{|d\widetilde{\theta}/dx|} = \frac{2}{3}\frac{\delta\widetilde{m}^2 (1+z)}{E_{\nu}\sin 4\widetilde{\theta}  (dz/dx)} \, ,
\end{equation}
where $dz/dx$ is given by the expansion rate of the universe. For $\delta m^2 \geq 10^{-17}\text{eV}^2$ and $E_{\nu} < 1$~PeV, we have $\gamma > 1$, which indicates that the evolution is adiabatic. In this adiabatic regime, the $\nu_\alpha\to \nu_\beta$ oscillation probability is given by
\begin{align}
P_{\alpha\beta} &= \sum_{j} |U_{\alpha j}|^2|U_{\beta j}|^2\left[\cos^2\widetilde{\theta}^{i}_{j}\cos^2\widetilde{\theta}^{f}_{j} + \sin^2\widetilde{\theta}^{i}_{j}\sin^2\widetilde{\theta}^{f}_{j}\right.\nonumber\\ 
&\quad \left.+ \frac{1}{2} \sin 2\widetilde{\theta}^{i}_{j}\sin2\widetilde{\theta}^{f}_{j}\cos\left( \int dx \frac{\delta\widetilde{m}^2_{j}}{4E_{\nu}} \right) \right] \, , 
\label{eq:18}
\end{align}
where $\widetilde{\theta}^{i}$ and $\widetilde{\theta}^f$ correspond to the effective mixing angles [cf.~Eq.~\eqref{eq:9}] when the neutrinos were created and today, respectively. When the matter effect is small, $\widetilde{\theta}^i_j\simeq \widetilde{\theta}^f_j \simeq \widetilde{\theta}_j$ and $\delta \widetilde{m}_j^2$ can be taken out of the integral. In this case, Eq.~\eqref{eq:18} simply reduces to Eq.~\eqref{eq:15}. In the parameter regime where the adiabaticity condition is not satisfied, we cannot express the oscillation probability analytically as in Eq.~\eqref{eq:18}. In such cases, we compute the oscillation probability purely numerically from the solution of the evolution equation, i.e.~$\langle \nu_\beta|\nu_\alpha\rangle(t) = \exp[-i\int_0^t dt' H (t')]$.

Note that in Eq.~\eqref{eq:18}, the effective oscillation length, as well as $\delta \widetilde m_j^2$, is now a function of the redshift. In particular, for the active-sterile oscillations to take effect, the oscillation length  $L_{\rm osc}$ must be comparable to or smaller than the effective source distance, given by~\cite{Carloni:2022cqz}
\begin{align}
    L_{\rm eff} = \int_{z_{\rm min}}^{z_{\rm max}} \frac{c \: dz}{H(z)(1+z)^2} \, ,
    \label{eq:Leff}
\end{align}
where the Hubble parameter is 
\begin{align}
H(z)=H_0\sqrt{\Omega_{\rm m}(1+z)^3+\Omega_\Lambda+(1-\Omega_{\rm m}-\Omega_\Lambda)(1+z)^2} \, ,
\end{align}
where $\Omega_{\rm m}$ and $\Omega_\Lambda$ are the fractions of matter (both visible and dark) and dark energy content in the Universe, respectively. We use the best-fit values from Planck data: $\Omega_{\rm m}=0.315$,  $\Omega_\Lambda=0.685$ and $H_0=67.4~{\rm km}\cdot{\rm s}^{-1}\cdot{\rm Mpc}^{-1}$~\cite{Planck:2018vyg}. Because of this choice of the unit for $H_0$, we have shown the speed of light $c$ explicitly in Eq.~\eqref{eq:Leff} to make it dimensionally correct. 
As for the maximum redshift value, we will take $z_{\rm max}=5$, beyond which the star formation rate decreases rapidly~\cite{Hopkins:2006bw, Yuksel:2008cu}, and  we do not expect any astrophysical sources of high-energy neutrinos to exist beyond this redshift. Similarly, for the minimum redshift, we take $z_{\rm min}=10^{-7}$, corresponding to the galactic center. 
Since the galactic contribution to the high-energy neutrino flux at IceCube is sub-dominant~\cite{IceCube:2023ame}, taking even smaller values of $z_{\rm min}$ will not significantly affect our results.

\section{Including C$\nu$B Overdensity}
\label{sec:over}

\begin{figure}[t!]
    \centering    \includegraphics[width=0.49\textwidth]{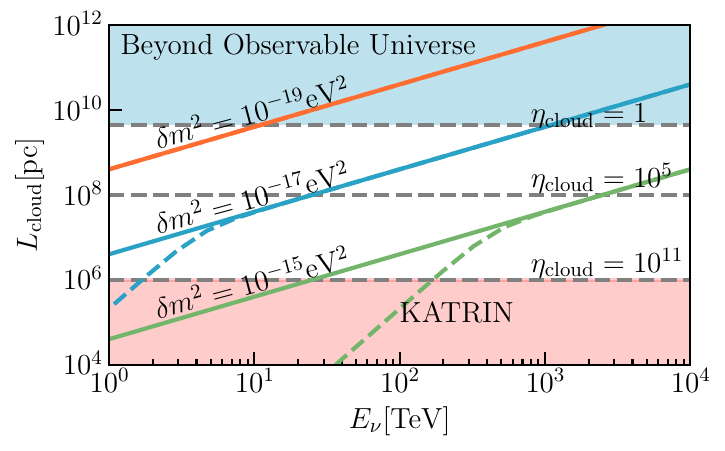}  
    \caption{The active-sterile mass splittings in vacuum ($\delta m^2$, solid) and including C$\nu$B matter effect ($\delta \widetilde{m}^2$, dashed) as a function of the neutrino energy. 
    The vertical axis shows the maximum size of the overdense C$\nu$B cluster for a given value of overdensity $\eta_{\rm cloud}$, demanding that the total number of relic neutrinos in the Universe is constant. 
    The red-shaded region at the bottom corresponds to the KATRIN exclusion limit on $\eta<1.1\times 10^{11}$ at 95\% CL~\cite{KATRIN:2022kkv}. 
    The blue-shaded region at the top corresponds to oscillation lengths beyond the observable Universe. }  
\label{fig:over}
\end{figure}

The values of $\delta m^2$ that are sensitive to the C$\nu$B matter effect very much depend on the incoming energy of the high-energy neutrinos. This is illustrated in Fig.~\ref{fig:over} by the solid lines for three benchmark values of $\delta m^2$. The corresponding dashed lines show the fixed $\delta \widetilde{m}^2$ values. The deviation of the dashed lines from the solid lines, therefore, represent the size of the matter effect. As we will see below, the C$\nu$B matter effect on the oscillation probabilities turns out to be negligible for the $\Lambda$CDM value of the C$\nu$B number density [cf.~Eq.~\eqref{eq:11}], especially for the $\delta m^2$ values required for adiabatic evolution.  Therefore, we allow for the possibility that there might be a local overdensity of C$\nu$B, parameterized by the ratio $\eta=n_\nu/n_{\nu,0}(1+z)^3$, i.e. the potential in Eq.~\eqref{eq:pot} is enhanced by a factor of $\eta$.\footnote{This is consistent with the cosmological constraints on the neutrino asymmetry, assuming that the clustering happens at late times, because $\eta_{\nu_\alpha}$ [cf.~Eq.~\eqref{eq:eta}] is constrained mainly from the BBN/CMB data.} The current experimental limit on $\eta$ is rather loose, only at the level of $10^{11}$ from KATRIN~\cite{KATRIN:2022kkv}, as shown by the red-shaded region in Fig.~\ref{fig:over}. See Refs.~\cite{Brdar:2022kpu, Bauer:2022lri, Tsai:2022jnv,Ciscar-Monsalvatje:2024tvm, Franklin:2024enc, DeMarchi:2024zer, Herrera:2024upj} for other local and global constraints on $\eta$, as well as future prospects. We assume a local overdensity around the Earth so that the matter effect is isotropic. Theoretically, while gravitational clustering alone can only give an ${\cal O}(1)$ enhancement~\cite{Ringwald:2004np, Mertsch:2019qjv, Zimmer:2023jbb, Elbers:2023mdr, Holm:2024zpr}, possible nonstandard neutrino (self)interactions that would cluster the C$\nu$B more than
gravity could in principle give $\eta\gg 1$, while avoiding the constraint from the Pauli exclusion principle~\cite{Bondarenko:2023ukx}. For instance, in a model with attractive Yukawa interactions mediated by an ultralight scalar, neutrino-antineutrino pairs can form stable clusters with $\eta_{\rm max}\sim 10^7$~\cite{Smirnov:2022sfo}. Without resorting to any particular new physics model, we just show a few benchmark values of $\eta$ in Fig.~\ref{fig:over} to illustrate our point. Note that for smaller $\eta$ values, the $\delta m^2$ and $\delta \widetilde{m}^2$ contours are identical, i.e. the matter effect is negligible. However, for $\eta\gtrsim 10^5$, we start to see the deviation of $\delta \widetilde{m}$ from $\delta m^2$, which implies that the matter effect is non-negligible.

An important thing to keep in mind is that, for a fixed number of total relic neutrinos in the Universe, $\eta> 1$ would imply that there is a maximum size for the overdense cluster, $L_{\rm cloud}=(c/H_0)\eta^{-1/3}$ assuming a spherical cluster with uniform density.\footnote{Other types of density profiles, such as the Einasto profile~\cite{1969Ap......5...67E}, would introduce more unknown parameters, like the spectral index and effective radius.}  
This is to ensure that the relic neutrinos do not overclose the Universe. 
For instance, as shown in Fig.~\ref{fig:over}, $\eta=1$ corresponds to $L_{\rm cloud}=c/H_0\simeq 4.5$ Gpc, which is roughly the size of the observable Universe, whereas $\eta=10^5$ corresponds to $L_{\rm cloud}\simeq 96$ Mpc, and $\eta=10^{11}$ corresponds to $L_{\rm cloud}\simeq 0.96$ Mpc. 
Therefore, for the matter effect to be relevant, we must have the effective oscillation length $L_{\rm osc}$ [cf.~Eq.~\eqref{eq:Losc}] comparable to or smaller than $L_{\rm cloud}$. We should clarify here that in principle, one does not need to saturate the upper bound on $L_{\rm cloud}$. For example, there could be multiple clusters each with smaller overdensities than in a single overdense cluster. For the mechanism presented in Ref.~\cite{Smirnov:2022sfo}, the size of each cluster can vary widely from $\sim$ km to $\sim 5$ Mpc. Since we do not assume any particular model for clustering, we simply use the maximum allowed value of $L_{\rm cloud}$ for concreteness. 

This in turn dictates the minimum value of $\delta m^2$ (for a given $E_\nu$), or the maximum value of $E_\nu$ (for a given $\delta m^2$), at which  the matter effect starts becoming important. For example, for $\delta m^2=10^{-19}~{\rm eV}^2$, the matter effect is not important in the entire energy range shown in Fig.~\ref{fig:over}, whereas for $\delta m^2=10^{-17}~{\rm eV}^2$, it starts becoming important for $E_\nu<70$ TeV, and for $\delta m^2=10^{-15}~{\rm eV}^2$, it is important for $E_\nu<70$ PeV, i.e. almost in the entire IceCube energy range of interest. However, this does not necessarily mean that IceCube has better sensitivity for higher $\delta m^2$ values, as this will depend on the actual oscillation probabilities, which we will discuss in Section~\ref{sec:results}.

For $\eta>1$, or a finite $L_{\rm cloud}<c/H_0$, we have to consider the case where the neutrinos were emitted from the distant source at an early redshift ($z_i\leq z_{\rm max}$)  and, after traveling through vacuum for some distance, encounter the C$\nu$B overdensity cloud at a redshift $z_c<z_i$ that creates a matter potential for them. 
In this case, the oscillation probability contains two parts: (i) vacuum probability from redshift $z_c$ to $z_i$, and (ii) matter probability from redshift $z_{\rm min}$ and $z_c$. 
Thus, Eq.~\eqref{eq:18} is modified to  
\begin{figure*}[t!]
    \centering    
    \includegraphics[width=0.49\textwidth]{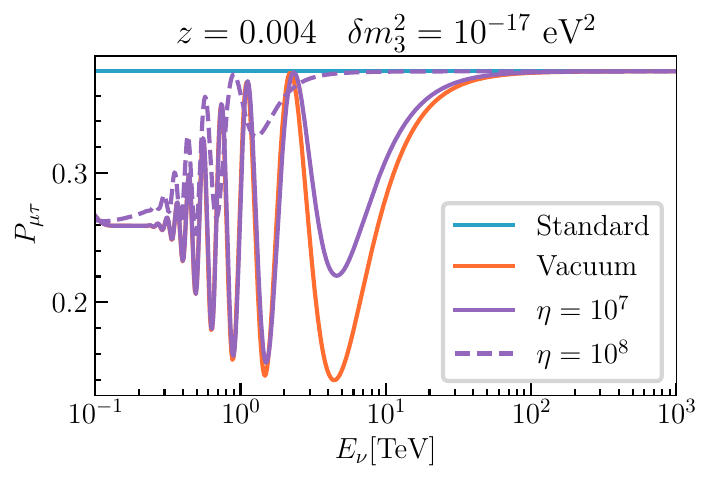}
    \includegraphics[width=0.49\textwidth]{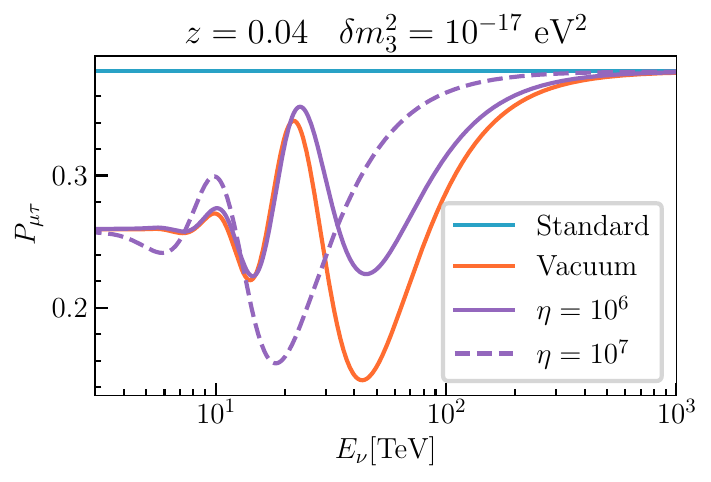}
    \caption{$\mu\to \tau$ oscillation probability as a function of neutrino energy for the pseudo-Dirac case in vacuum only (orange) and including the C$\nu$B matter effect (purple) with two benchmark values of $\eta$ (solid and dashed). Here we have fixed $\delta m_3^2=10^{-17}~{\rm eV}^2$ and $z=0.004$ (0.04) in the left (right) panel. The size of the local overdensity $L_{\text{cloud}}$ extends up to 10 Mpc, 21 Mpc and 45 Mpc for $\eta=10^8$, $10^7$ and $10^6$, respectively. An asymmetry of 1\% between neutrinos and antineutrinos is assumed. The fast oscillations at low energies are averaged out. The averaged oscillation probability in the standard 3-neutrino case (blue) is also shown for comparison. }
\label{prob}
\end{figure*}
%
\begin{align}
P_{\alpha\beta} &= \frac{1}{2}\sum_{j} |U_{\alpha j}|^2|U_{\beta j}|^2
\left[1 + \cos 2\widetilde{\theta}^{i}_{j}\cos 2\widetilde{\theta}^{f}_{j}\cos\left(\frac{\delta m^2_{j}L_{\rm eff}}{4E_{\nu}}\right)\right. \nonumber \\ 
& \left.
+ \sin 2\widetilde{\theta}^{i}_{j}\sin2\widetilde{\theta}^{f}_{j}\cos\left( \int dx \frac{\delta\widetilde{m}^2_{j}}{4E_{\nu}} + \frac{\delta m^2_{j}L_{\rm eff}}{4E_{\nu}} \right) \right].  
\label{eq:22}
\end{align}
%
Notice that in this case, $\widetilde{\theta}^{i}$ and $\widetilde{\theta}^f$ correspond to the effective mixing angles when the neutrinos arrive to the C$\nu$B cloud and today, respectively. Also, $L_{\rm eff}$ is given by Eq.~\eqref{eq:Leff} but with the lower limit of integration replaced by $z_c$, which is the redshift distance equivalent of $L_{\rm cloud}$. Basically, in vacuum, we can take $\delta m^2$ out of the redshift integral, whereas in matter, we have to keep $\delta \widetilde{m}^2$ inside the integral, since it also depends on the redshift. For $\eta\lesssim 10^4$, when the matter effect is negligible, the last two contributions inside the parenthesis can be combined into one that exactly becomes equal to $\delta m^2_j L_{\rm eff}/4E_\nu$ as in the second term, and Eq.~\eqref{eq:22} simply reduces to the vacuum oscillation result [cf.~Eq.~\eqref{eq:18} with tildes removed].

\section{Results} \label{sec:results}

To understand the energy dependence of the oscillation probabilities in the presence of matter effect, we plot the $\nu_\mu\to \nu_\tau$ oscillation probabilities\footnote{Similar behavior is observed for other flavors, and therefore, we do not show all of them here.} for the standard and pseudo-Dirac cases (with and without matter effect) in Fig.~\ref{prob}. 
Here we have fixed the active-sterile mass splitting for just one pair: $\delta m_3^2=10^{-17}~{\rm eV}^2$, while keeping $\delta m_1^2=\delta m_2^2=0$. 
In the left panel, we have fixed the source redshift distance at $z=0.004$, which is roughly the distance to NGC 1068, the most significant point source identified by IceCube~\cite{IceCube:2022der}. The vacuum oscillation probability for the pseudo-Dirac case is noticeably different from the standard case for $E_\nu\lesssim 50$ TeV. At higher energies, the effective oscillation length exceeds the source distance, and therefore, the vacuum oscillation probability approaches the standard case. On the other hand, at low energies, the fast oscillations are averaged out to a constant value (but different from the standard case). Now including the matter effect further modifies the oscillation probability, as it tends to suppress the oscillation amplitude, as compared to the vacuum case. But this effect is observable only for $\eta\gg 1$, because the source is relatively nearby, so we need a large $\eta$ to be able to make a significant contribution to the third term in Eq.~\eqref{eq:22}. Here we have chosen two benchmark values of $\eta=10^7$ and $\eta=10^8$, corresponding to $L_{\rm cloud}=21$ Mpc and 10 Mpc, respectively.  As we increase the size of the matter effect by cranking up $\eta$, the oscillation extrema are also shifted to lower energies.

In the right panel of Fig.~\ref{prob}, we keep the same $\delta m_3^2=10^{-17}~{\rm eV}^2$, but increase the source distance to $z=0.04$. In this case, the oscillations are shifted to higher energies, and the pseudo-Dirac oscillations are noticeably different from the standard one for $E_\nu\lesssim 500$ TeV. Also, since the source is further away, a slightly smaller value of $\eta=10^6$ corresponding to $L_{\rm cloud}=45$ Mpc is now sufficient to induce a noticeable matter effect. As in the left panel, increasing $\eta$ shifts the oscillation extrema to lower energies, before they approach the fast oscillations. Since getting very large $\eta$ values is theoretically challenging, we will fix a benchmark value of $\eta=10^6$ and $z=0.04$ for the flavor triangle analysis below.  

For a given source distance, if we increase the $\delta m^2$ value, the oscillations will also be shifted to higher energies. Since the astrophysical neutrino flux is expected to have a power-law behavior~\cite{Fang:2022trf}, going to higher energy means having smaller flux, and hence, less statistics. It turns out that IceCube will eventually lose sensitivity for $\delta m^2\gtrsim 10^{-16}~{\rm eV}^2$~\cite{Carloni:2022cqz}. Therefore, we use $\delta m^2=10^{-17}~{\rm eV}^2$ as our benchmark value.

\begin{figure}[t!]
    \centering    
    \includegraphics[width=0.49\textwidth]{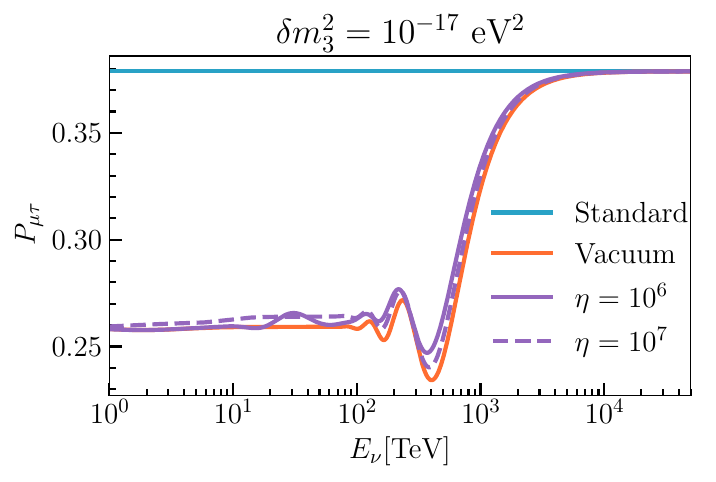}
    \caption{Same as in Fig.~\ref{prob}, but here we have averaged over the distance traveled by the neutrinos and have assumed a flat distribution of sources up to $z=5$. }
\label{prob2}
\end{figure}

In Fig.~\ref{prob2}, we have plotted the same $\nu_\mu\to \nu_\tau$ probabilities for the standard and pseudo-Dirac (vacuum and matter) cases as a function of energy, but here we have averaged over the the distances up to redshift $z_{
\rm max}=5$, assuming a flat distribution of sources. The first dip in the probability at the highest energy is due to contributions from sources at $z=5$. 
We note that increasing $\eta$ (or decreasing the cloud size) makes this dip closer to the vacuum case because the neutrinos mostly travel in vacuum;  therefore, going to an arbitrarily high overdensity is actually not helpful for disentangling the matter effect.
As we go to lower energies, the sources at smaller redshifts cause multiple oscillations, which eventually average out and approach the vacuum result, as also noted in Fig.~\ref{prob}. 
On the other hand, both vacuum and matter oscillations approach the standard result at energies beyond 5 PeV, since the effective oscillation length for the chosen mass splitting goes beyond $z=5$. If the sources do not follow a flat distribution, as might be the case if they follow the star formation rate~\cite{Yuksel:2008cu}, the features observed in Fig.~\ref{prob2} will shift to lower energies. 

\begin{figure*}[t!]
    \centering   
    \includegraphics[width=0.49\textwidth]{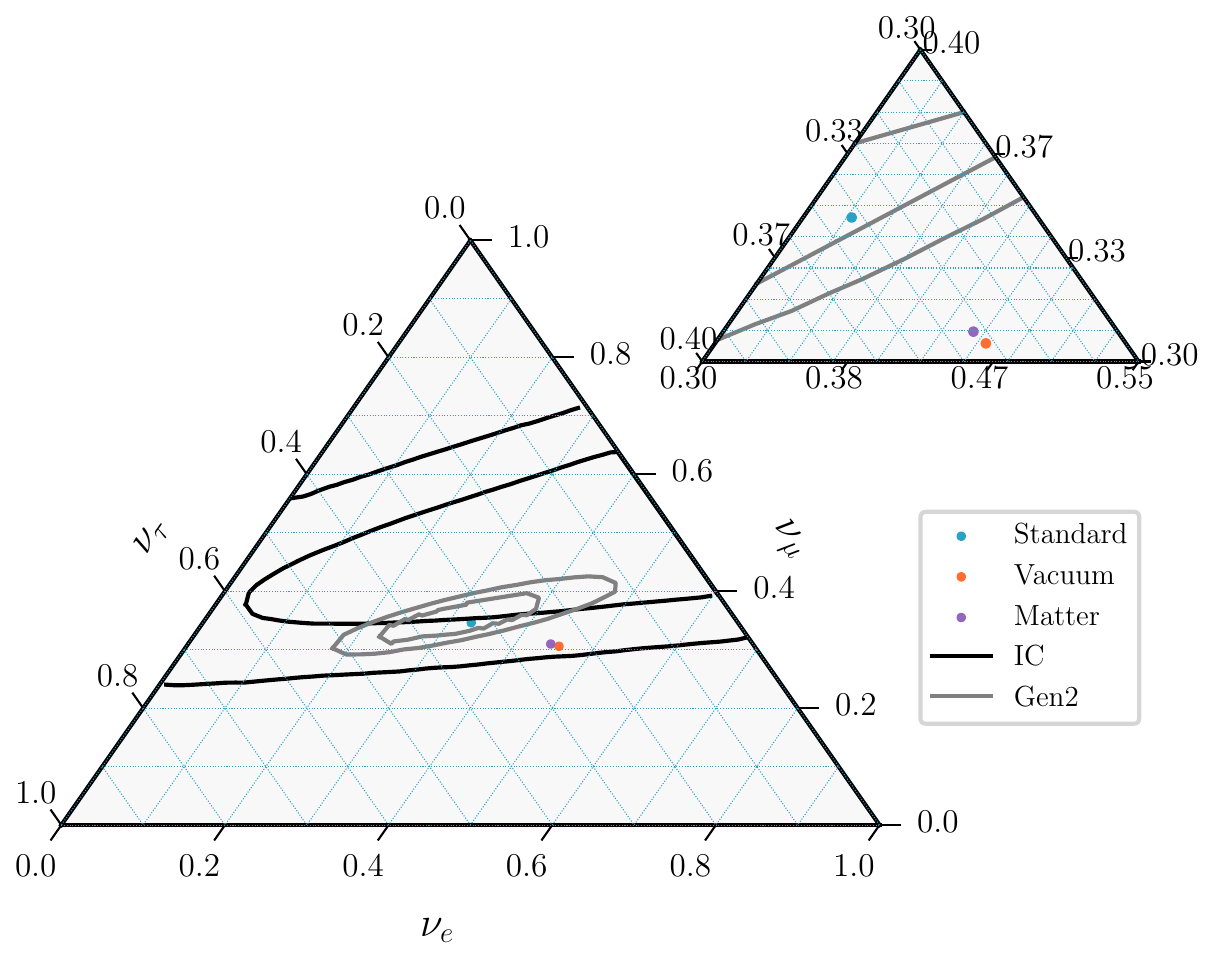}
     \includegraphics[width=0.49\textwidth]{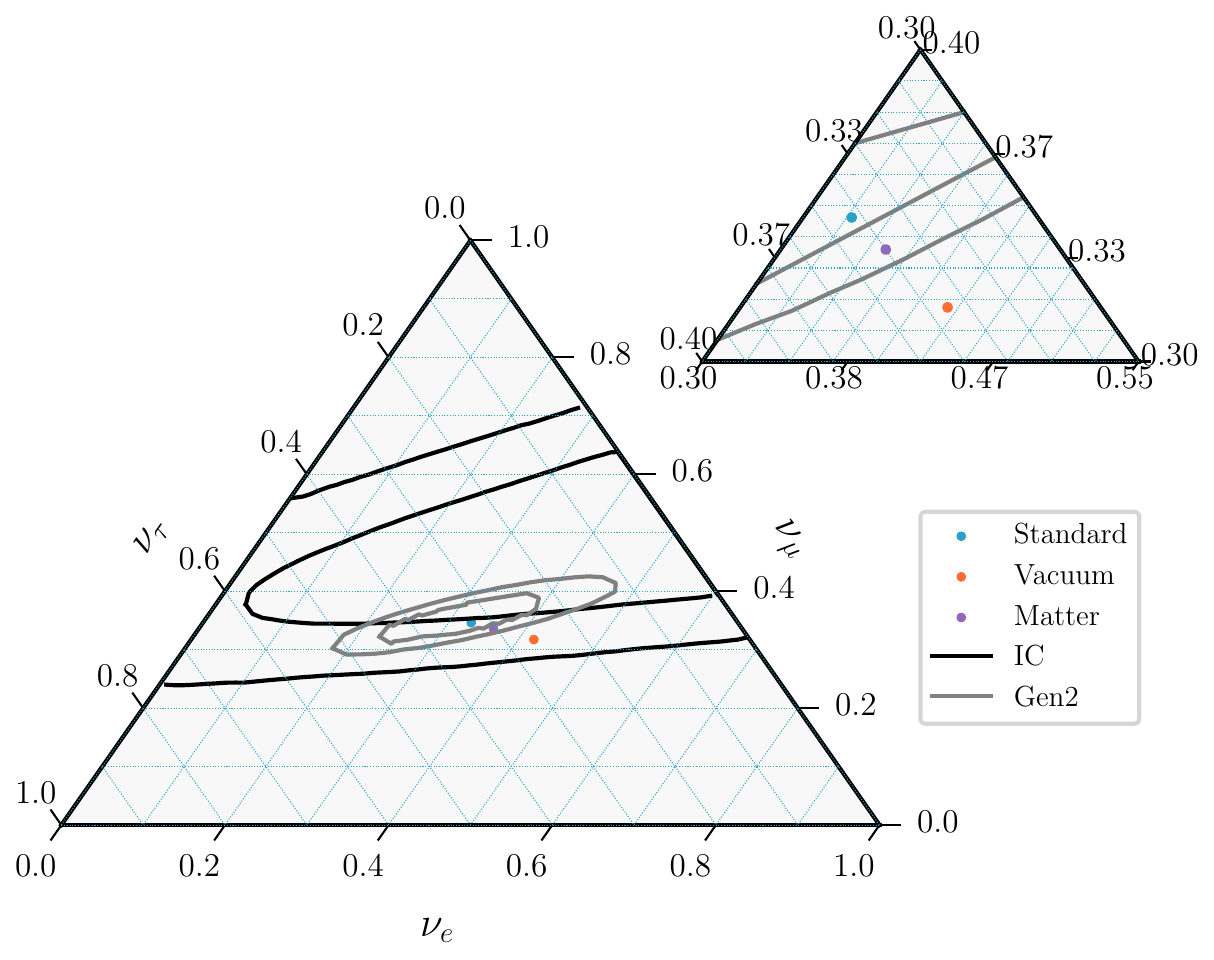}
    \caption{C$\nu$B matter effect on the best-fit point in the flavor triangle for $E_\nu=1$ TeV (left) and 40 TeV (right). With increasing energy, the best-fit point for the pseudo-Dirac case in the presence of matter effect (purple) moves from the vacuum case (orange) toward the standard 3-neutrino case (light blue). Here we have fixed $\delta m_3^2=10^{-17}~{\rm eV}^2$, $z=0.04$ and have considered the standard pion source with initial flavor ratio $(1/3:2/3:0)$. For the matter case, we have fixed $\eta=10^6$ that corresponds to an overdensity size of $L_{\text{cloud}}=45$~Mpc. An asymmetry of 1\% it is assumed}. 
\label{ternary1}
\end{figure*}

After calculating the effect of pseudo-Dirac neutrino oscillations on the probabilities, we are now in a position to compare the final flavor ratio results for the standard and pseudo-Dirac case with and without matter effect. This is shown in  Fig.~\ref{ternary1}. 
Note that it is important to compare only the normalized flavor ratios, because the total flux of active neutrinos in the pseudo-Dirac case may not be conserved due to active-sterile oscillations; therefore, $\sum_\beta f_{\beta,\oplus}$ calculated from Eq.~\eqref{eq:fstd} is not necessarily guaranteed to be unity for the pseudo-Dirac case, unlike in the standard case.  
Here we take a standard pion decay source: 
\begin{align}
    \pi^\pm \to \mu^\pm + \overset{\scriptscriptstyle(-)}{\nu_\mu}\to e^\pm+\overset{\scriptscriptstyle(-)}{\nu_e}+\nu_\mu+\bar{\nu}_\mu \, ,
\end{align}
with an initial flavor composition of  $(1/3:2/3:0)$. 
With the best-fit values for the oscillation parameters taken from {\tt NuFit}~\cite{nufit} and assuming a normal mass ordering, the standard 3-neutrino vacuum oscillation paradigm predicts a final flavor ratio of $(0.33:0.35:0.32)$, as shown by the blue dot. 
On the other hand, for our benchmark pseudo-Dirac case with $\delta m_3^2=10^{-17}~{\rm eV}^2$, just considering vacuum oscillations from a source at redshift $z=0.04$ gives us $(0.46 : 0.30: 0.24)$ at $E_\nu=1$ TeV and $(0.42:0.32:0.26)$ at $E_\nu=40$ TeV, as shown by the orange dots in the two panels. Note the mild energy-dependence of the best-fit value here. This was also noted in Ref.~\cite{Shoemaker:2015qul}. 

Including the C$\nu$B matter effect for an overdensity of $\eta=10^4$ makes the energy-dependent effects more prominent in the oscillation probabilities (see Figs.~\ref{prob} and \ref{prob2}). 
In the left panel, we show the result for $E_\nu=1$ TeV, where the matter effect gives a best-fit flavor ratio of $(0.44:0.31:0.25)$, while in the right panel with $E_\nu=40$ TeV, it gives $(0.36:0.34:0.30)$. Thus, as we go from lower to higher energies, the best-fit point moves from the vacuum case to the standard case, as can be clearly seen from the inset plots. 
The energy window of TeV-PeV is optimal for this effect to be observable. 
For very high energies, the neutrino flux goes down rapidly and the event statistics will be low. 
On the other hand, for energies smaller than a few TeV, the atmospheric background will be overwhelming. 
Moreover, the tau neutrinos are not detectable at IceCube for low energies; the lowest-energy tau event observed so far is at 20 TeV~\cite{IceCube:2024nhk}.   

In Fig.~\ref{ternary1}, the current 68\% and 90\% CL IceCube limits~\cite{IceCube:2015gsk}  are shown by the black contours.\footnote{
Preliminary tighter constraints are reported in Ref.~\cite{IceCube:2023fgt} by adding more years of data and updated ice properties on the HESE sample, but we show the officially published results from Ref.~\cite{IceCube:2015gsk}.} 
The future prospects for flavor triangle measurements are bright~\cite{Ackermann:2022rqc}, with the observation of high-energy neutrinos by several next-generation neutrino telescopes, such as IceCube-Gen 2~\cite{IceCube-Gen2:2020qha}, KM3NeT~\cite{KM3Net:2016zxf}, Baikal-GVD~\cite{Allakhverdyan:2021vkk},  P-ONE~\cite{P-ONE:2020ljt}, TRIDENT~\cite{Ye:2022vbk}, TAMBO~\cite{Thompson:2023pnl}, Trinity~\cite{Otte:2019aaf} and RET~\cite{RadarEchoTelescope:2021czz}. 
The possibility of a joint analysis of the combined data from multiple experiments sensitive to different neutrino flavors (e.g., cascade and track data from IceCube-Gen 2, combined with the tau-neutrino data from TAMBO) could significantly improve the precision on the flavor triangle data.    
For illustration, we show the IceCube-Gen 2 projections~\cite{IceCube-Gen2:2023rds} by the grey  contours. It is clear that while the current IceCube constraint is not enough to probe the C$\nu$B matter effect, the IceCube-Gen 2 will be able to do so. In fact, it can clearly distinguish the energy-dependent matter effect from the vacuum oscillations, which will provide a new way to probe the C$\nu$B overdensity, on top of probing the pseudo-Dirac hypothesis.

\begin{figure*}[t!]
    \centering    \includegraphics[width=0.49\textwidth]{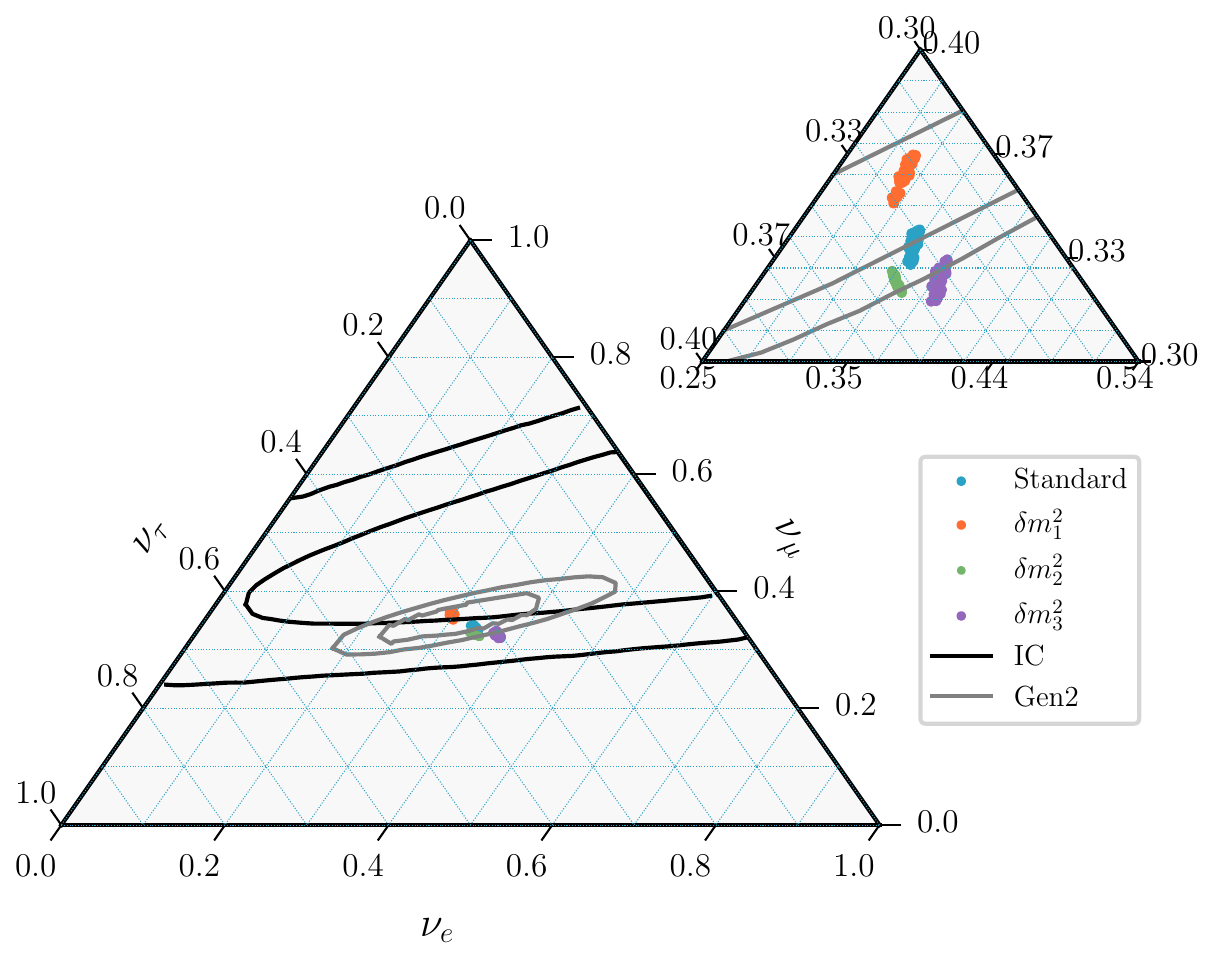}     \includegraphics[width=0.49\textwidth]{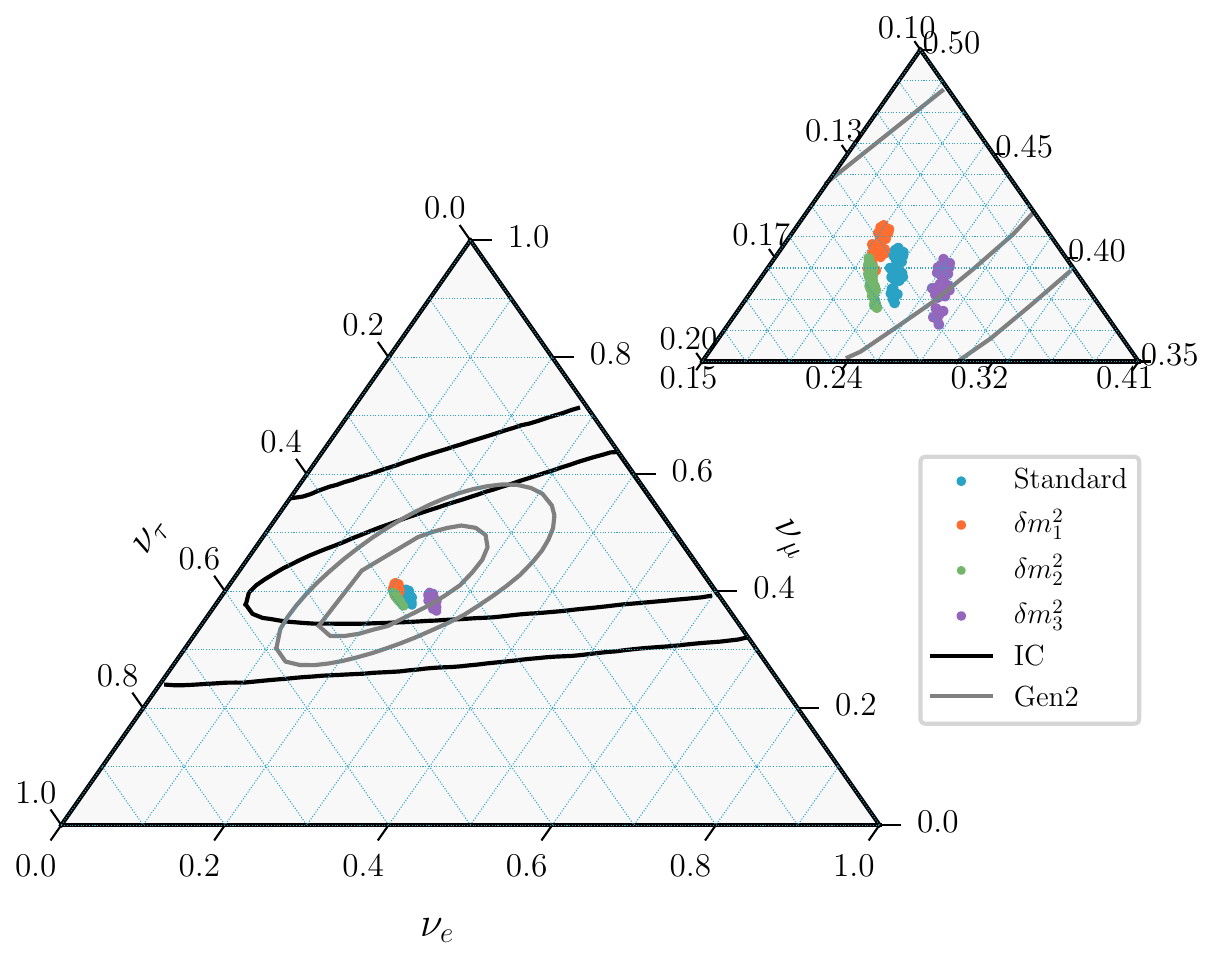}\\
    \includegraphics[width=0.49\textwidth]{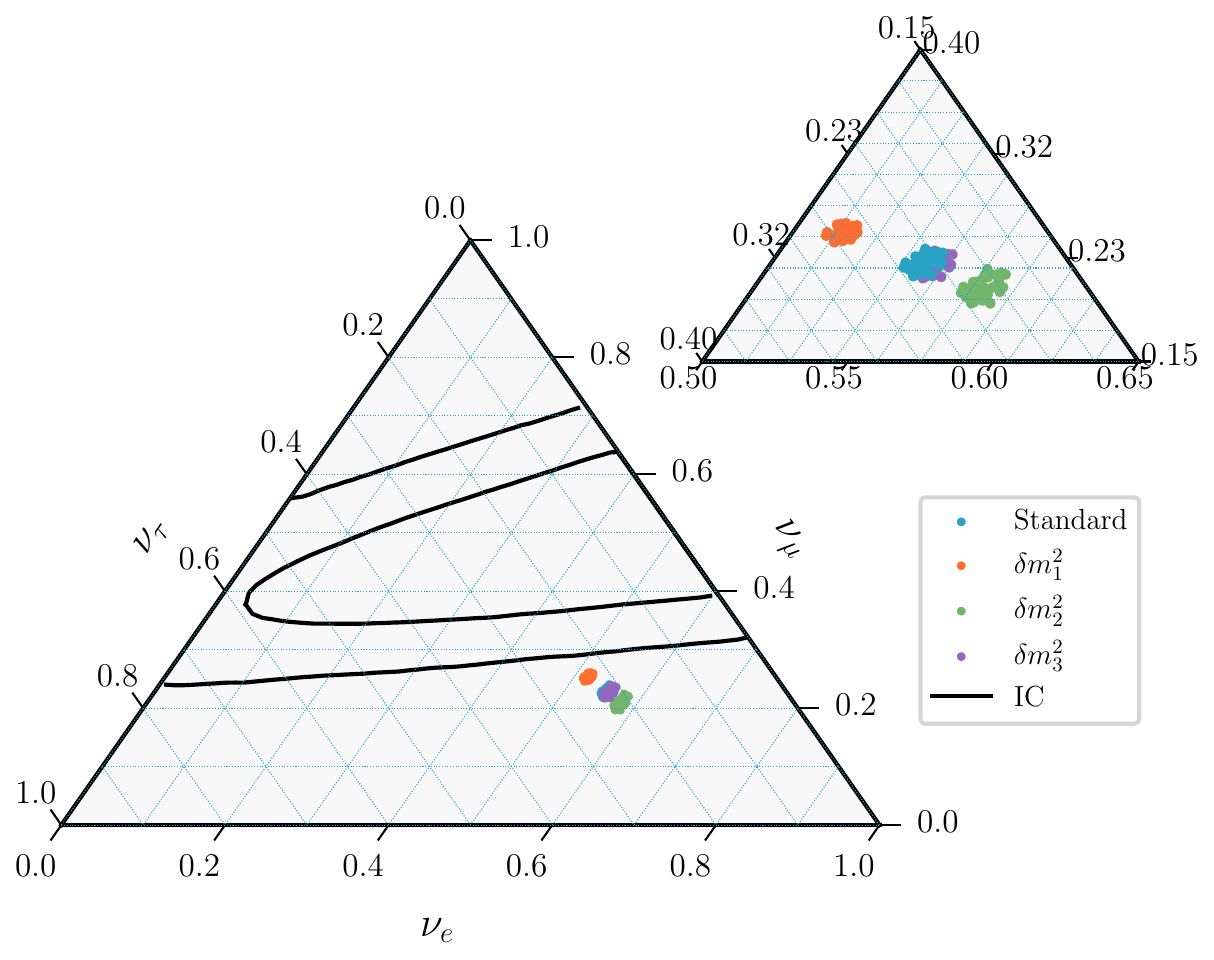}   \includegraphics[width=0.49\textwidth]{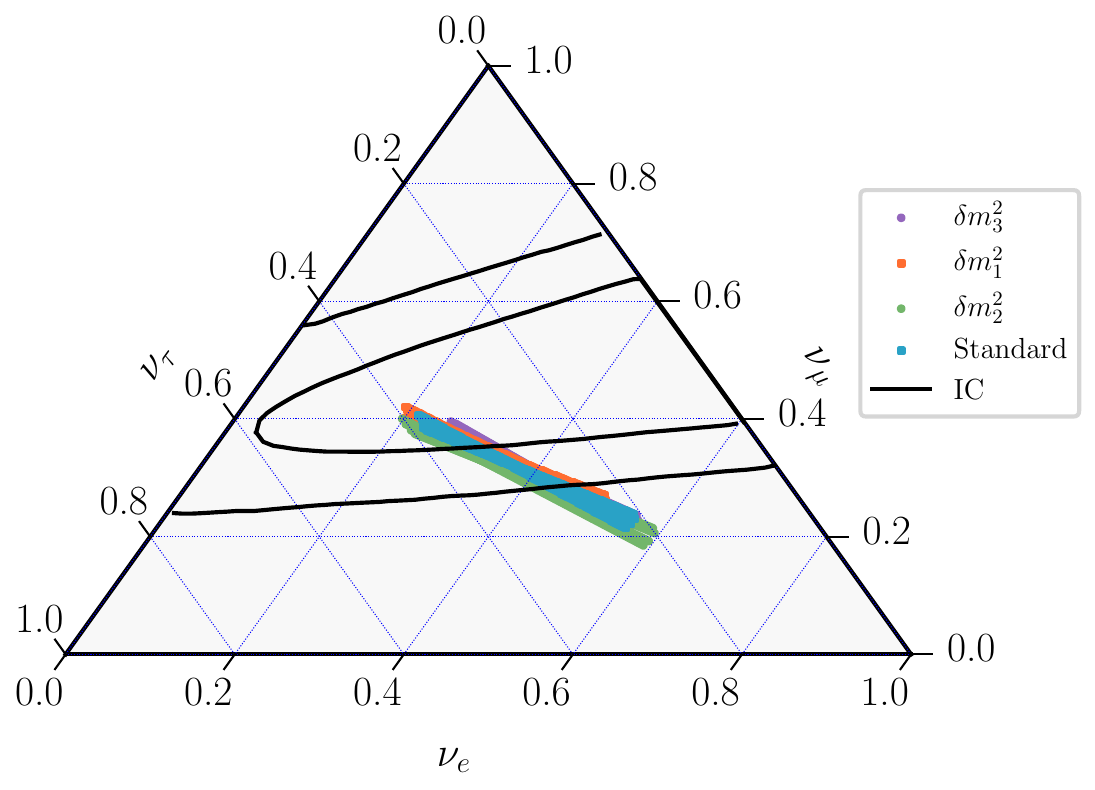}
    \caption{Ternary plots for the neutrino flavor composition on Earth for four different benchmark source flavor compositions (i) (1/3:2/3:0), (ii) (0:1:0),  (iii) (1:0:0), (iv) $(x:1-x:0)$. Here we compare the standard 3-neutrino oscillation paradigm with the pseudo-Dirac case with one active-sterile mass splitting nonzero. We have fixed the distance at $z=0.04$ and the neutrino energy at 40 TeV. For the matter case, we have fixed $\eta=10^6$ that corresponds to an overdensity size of $L_{\text{cloud}}=45$~Mpc}. 
\label{ternary2}
\end{figure*}

\begin{figure*}
    \centering    \includegraphics[width=0.49\textwidth]{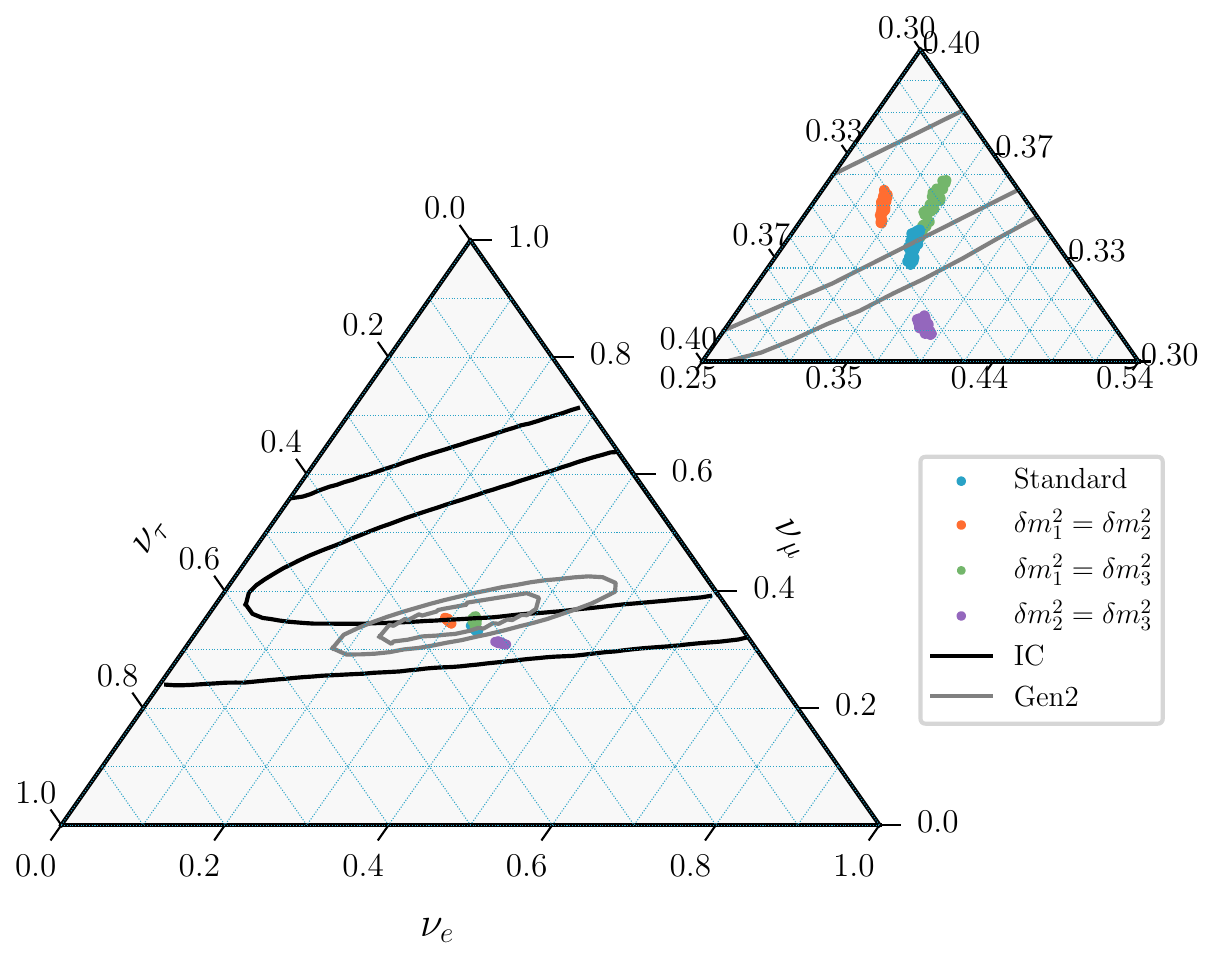}     \includegraphics[width=0.49\textwidth]{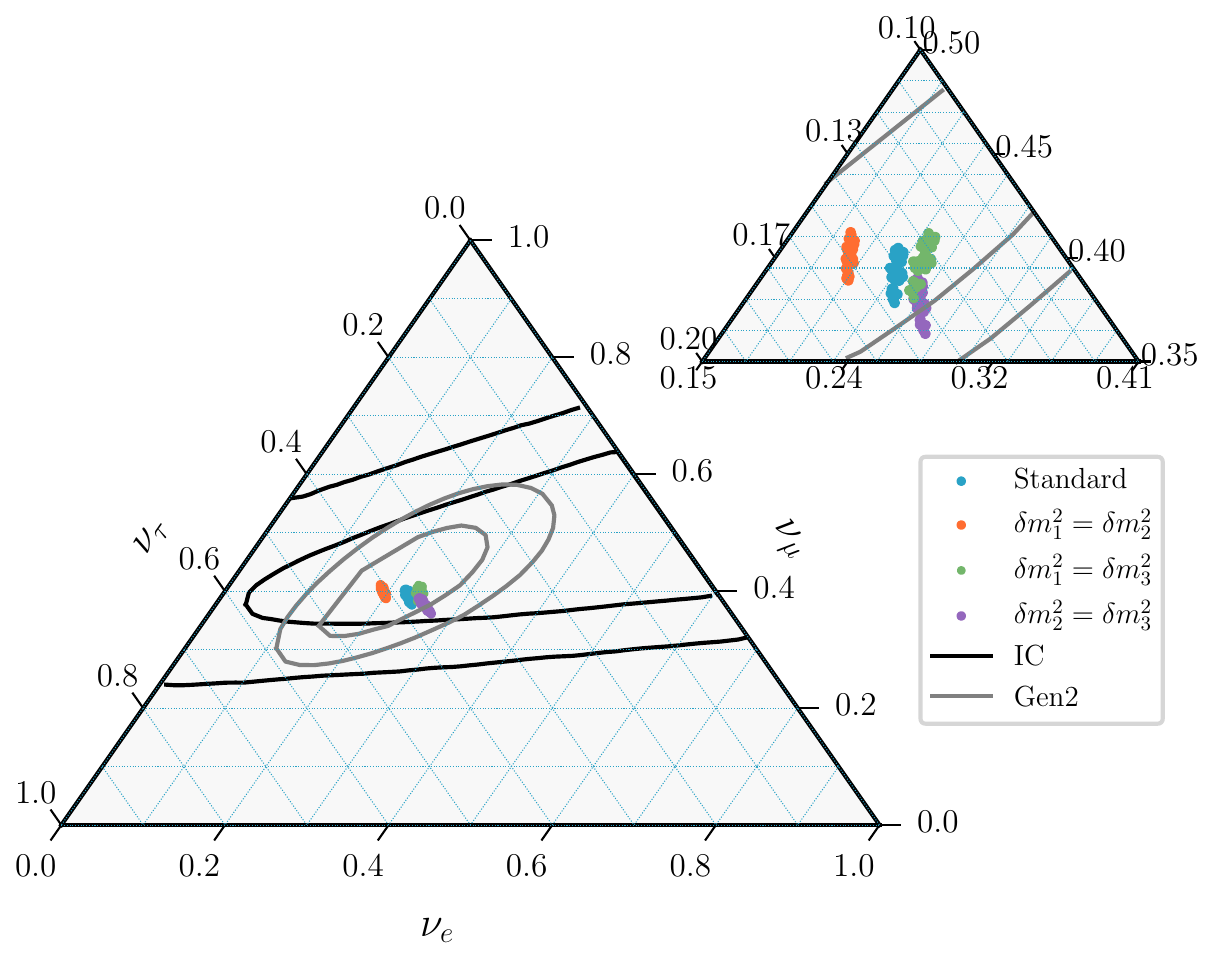}\\ \includegraphics[width=0.49\textwidth]{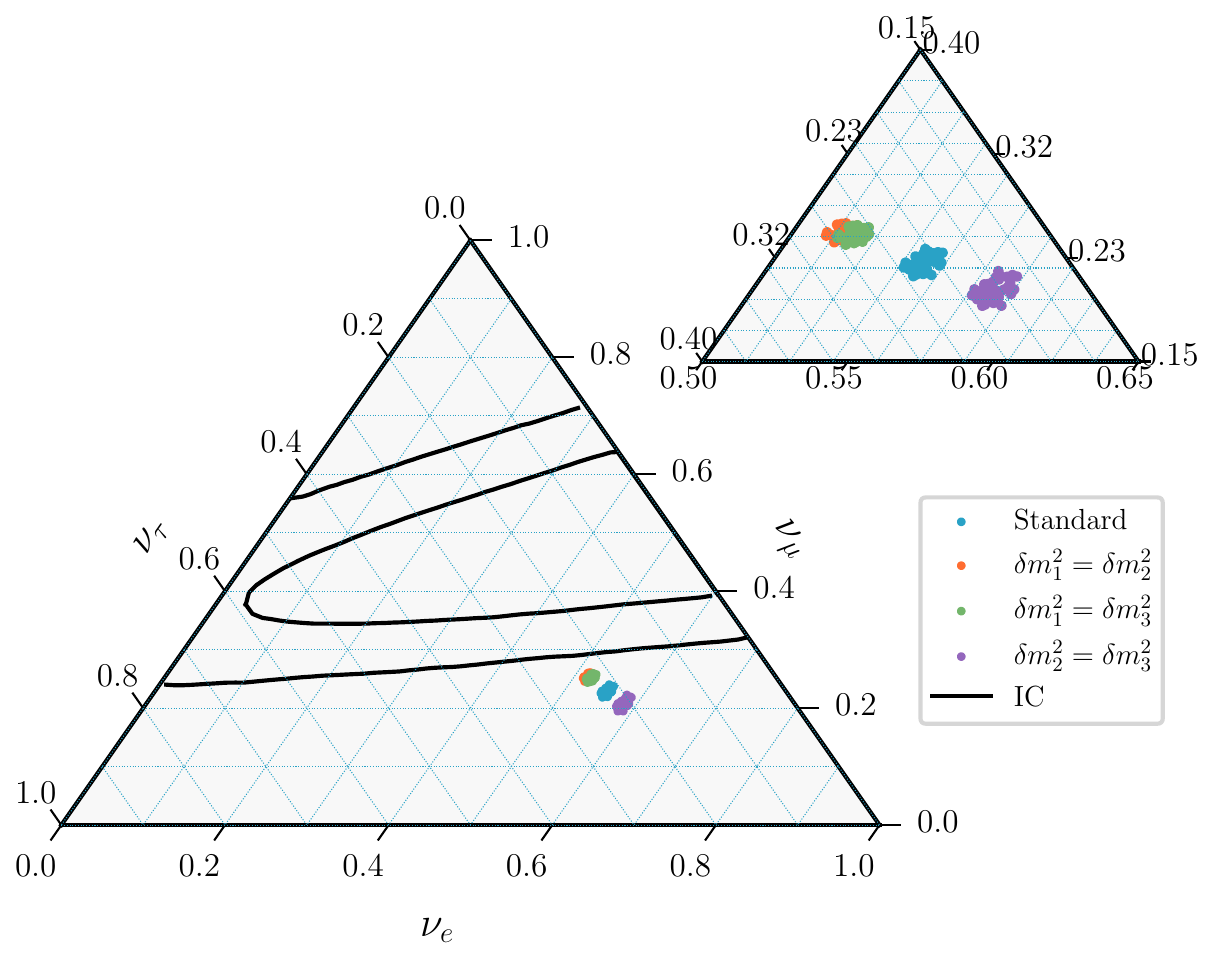}   \includegraphics[width=0.49\textwidth]{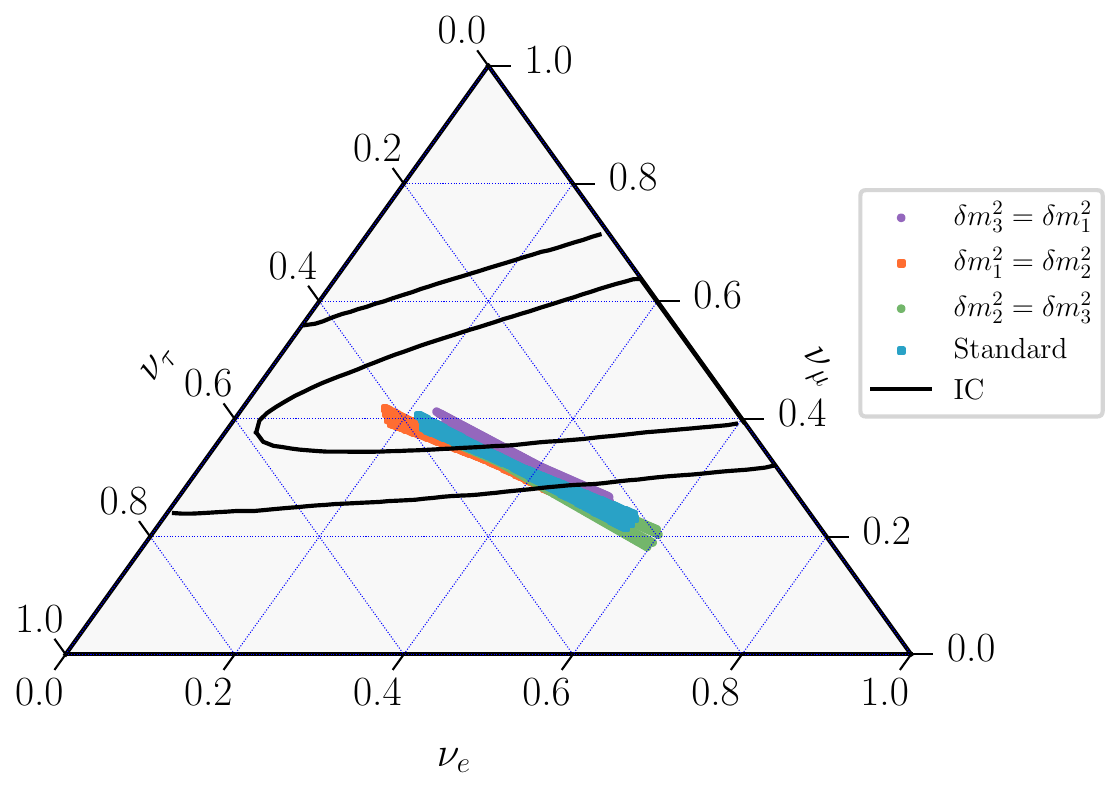}
    \caption{Same as Fig.~\ref{ternary1}, but with two active-sterile mass splittings nonzero. }
\label{ternary3}
\end{figure*}

In Fig.~\ref{ternary2}, we fix the energy at 40 TeV, but generalize our analysis to different initial flavor compositions, as mentioned in Section~\ref{sec:intro}, namely, (i) standard pion decay (top left panel), (ii) muon-suppressed pion decay (top right panel), (iii) neutron decay (bottom left panel), and (iv) general case (bottom right panel). 
We also include the variation of the mixing angles in their 68\% CL allowed range from {\tt NuFit}~\cite{nufit}, which results in a spread of the points for each case. 
Our use of the reduced uncertainties (68\% CL) is in anticipation of the precision measurements of the oscillation parameters at next-generation neutrino oscillation experiments, such as JUNO~\cite{JUNO:2015zny}, DUNE~\cite{DUNE:2015lol}, and Hyper-K~\cite{Hyper-Kamiokande:2018ofw}, before the next-generation neutrino telescopes start collecting data.
We find that with improved precision on the oscillation parameters, it is possible to distinguish the standard case from the pseudo-Dirac case for certain combinations of $\delta m^2_{i}$, assuming a known initial flavor composition at a specific energy.
The energy-dependent effect shown in Fig.~\ref{ternary1} will make this distinction even easier. 
We also notice that the separation from the standard case on the flavor triangle is different, depending on the initial flavor composition and on which $\delta m_i^2$ is nonzero. This information will provide a unique way to probe the individual active-sterile mass splittings in the pseudo-Dirac scenario. 

In Fig.~\ref{ternary3}, we further generalize our analysis to include two active-sterile mass splittings nonzero (but equal). 
Even in this case, the distinction between the standard and pseudo-Dirac cases, as well as between the different $\delta m_i^2$ pairs, can be made for a known initial flavor ratio. Of course, if the initial flavor composition is not known precisely, it becomes more difficult to distinguish the pseudo-Dirac case, as shown in the lower right panels of Figs.~\ref{ternary2} and \ref{ternary3}.  

Finally, when we have all three mass splittings nonzero and equal, their effect on the flavor ratio cancels out and there is no longer any difference with the standard case. However, we would still have reduction of the flux due to oscillation to sterile neutrinos, if the oscillation length is comparable to the distance traveled by the neutrinos, as this would produce oscillatory behavior in the energy spectrum of the neutrino flux. Thus, for certain range of mass splitting this reduction in flux can be energy dependent and therefore distinguishable from the standard case even without knowledge of the normalization of the initial flux.

\section{Conclusions}
\label{sec:con}
The flavor ratio measurements of the high-energy astrophysical neutrinos at neutrino telescopes provide crucial information on the source properties.  
We have shown that the flavor ratio predictions are altered from the standard 3-neutrino paradigm if neutrinos are pseudo-Dirac particles with tiny active-sterile mass splittings. 
In particular, we find for the first time that the C$\nu$B matter effect induces a novel energy-dependent flavor effect, which is robust against energy reconstruction, and hence, can be distinguished from other sources of energy dependence. 
We therefore advocate making energy-dependent flavor triangle measurements at neutrino telescopes. 
Energy-dependent flavor composition measurements were also advocated recently in Ref.~\cite{Liu:2023flr} for a different physics reason, i.e. to establish  the transition from neutrino production via the full pion decay chain at low energies to muon-damped pion decay at high energies. 
This is challenging today, but may be feasible in the future. 
Moreover, the matter effect strongly depends on the local C$\nu$B overdensity and the neutrino asymmetry, and therefore, a precise determination of the flavor composition at future neutrino telescopes can in principle provide a useful probe of the local C$\nu$B overdensity and neutrino asymmetry parameters.  This will be complementary to other existing probes of overdensity (e.g.~using cosmic ray upscattering~\cite{DeMarchi:2024zer}) as well as neutrino asymmetries (e.g. using BBN data~\cite{Matsumoto:2022tlr}). 

Beyond flavor measurements, the pseudo-Dirac scenario will induce oscillations in the neutrino energy spectrum. The oscillation length and amplitude depend on the local C$\nu$B overdensity. Thus, a precise measurement of neutrino energy could also contribute to probing the C$\nu$B.

\section*{Acknowledgments}
We thank Julien Froustey and Jack Shergold for useful discussions. We also thank the anonymous referee for pointing out an important error in the matter potential calculation. This work of BD was partly supported by the U.S. Department of Energy under grant No. DE-SC 0017987. PM is supported by Fermi Research Alliance, LLC under Contract No. DE-AC02-07CH11359 with the U.S. Department of Energy, Office of Science, Office of High Energy Physics. IMS is supported by STFC grant ST/T001011/1. BD and PM thank the organizers of the Mitchell Conference 2023, where this work was initiated. We also acknowledge the Center for Theoretical Underground Physics and Related Areas (CETUP*) and the Institute for Underground Science at SURF for hospitality and for providing a stimulating environment, where a part of this work was done.

{\bf Note Added:} While we were completing this work, Ref.~\cite{Fong:2024mqz} appeared, where the authors also discuss the effect of pseudo-Dirac neutrinos on the flavor triangle. 
But they have not included the C$\nu$B matter effect.

\bibliographystyle{utphys}
\bibliography{ref}

\providecommand{\href}[2]{#2}\begingroup\raggedright\begin{thebibliography}{100}

\bibitem{Wolfenstein:1981kw}
L.~Wolfenstein, ``{Different Varieties of Massive Dirac Neutrinos},''
  \href{http://dx.doi.org/10.1016/0550-3213(81)90096-1}{{\em Nucl. Phys. B}
  {\bfseries 186} (1981) 147--152}.

\bibitem{Petcov:1982ya}
S.~T. Petcov, ``{On Pseudodirac Neutrinos, Neutrino Oscillations and
  Neutrinoless Double beta Decay},''
  \href{http://dx.doi.org/10.1016/0370-2693(82)91246-1}{{\em Phys. Lett. B}
  {\bfseries 110} (1982) 245--249}.

\bibitem{Valle:1983dk}
J.~W.~F. Valle and M.~Singer, ``{Lepton Number Violation With Quasi Dirac
  Neutrinos},'' \href{http://dx.doi.org/10.1103/PhysRevD.28.540}{{\em Phys.
  Rev. D} {\bfseries 28} (1983) 540}.

\bibitem{Doi:1983wu}
M.~Doi, M.~Kenmoku, T.~Kotani, H.~Nishiura, and E.~Takasugi, ``{PSEUDODIRAC
  NEUTRINO},'' \href{http://dx.doi.org/10.1143/PTP.70.1331}{{\em Prog. Theor.
  Phys.} {\bfseries 70} (1983) 1331}.

\bibitem{Kobayashi:2000md}
M.~Kobayashi and C.~S. Lim, ``{Pseudo Dirac scenario for neutrino
  oscillations},'' \href{http://dx.doi.org/10.1103/PhysRevD.64.013003}{{\em
  Phys. Rev. D} {\bfseries 64} (2001) 013003},
  \href{http://arxiv.org/abs/hep-ph/0012266}{{\ttfamily arXiv:hep-ph/0012266}}.

\bibitem{Chang:1999pb}
D.~Chang and O.~C.~W. Kong, ``{Pseudo-Dirac neutrinos},''
  \href{http://dx.doi.org/10.1016/S0370-2693(00)00228-8}{{\em Phys. Lett. B}
  {\bfseries 477} (2000) 416--423},
  \href{http://arxiv.org/abs/hep-ph/9912268}{{\ttfamily arXiv:hep-ph/9912268}}.

\bibitem{Nir:2000xn}
Y.~Nir, ``{PseudoDirac solar neutrinos},''
  \href{http://dx.doi.org/10.1088/1126-6708/2000/06/039}{{\em JHEP} {\bfseries
  06} (2000) 039}, \href{http://arxiv.org/abs/hep-ph/0002168}{{\ttfamily
  arXiv:hep-ph/0002168}}.

\bibitem{Joshipura:2000ts}
A.~S. Joshipura and S.~D. Rindani, ``{Phenomenology of pseudoDirac
  neutrinos},'' \href{http://dx.doi.org/10.1016/S0370-2693(00)01148-5}{{\em
  Phys. Lett. B} {\bfseries 494} (2000) 114--123},
  \href{http://arxiv.org/abs/hep-ph/0007334}{{\ttfamily arXiv:hep-ph/0007334}}.

\bibitem{Lindner:2001hr}
M.~Lindner, T.~Ohlsson, and G.~Seidl, ``{Seesaw mechanisms for Dirac and
  Majorana neutrino masses},''
  \href{http://dx.doi.org/10.1103/PhysRevD.65.053014}{{\em Phys. Rev. D}
  {\bfseries 65} (2002) 053014},
  \href{http://arxiv.org/abs/hep-ph/0109264}{{\ttfamily arXiv:hep-ph/0109264}}.

\bibitem{Balaji:2001fi}
K.~R.~S. Balaji, A.~Kalliomaki, and J.~Maalampi, ``{Revisiting pseudoDirac
  neutrinos},'' \href{http://dx.doi.org/10.1016/S0370-2693(01)01356-9}{{\em
  Phys. Lett. B} {\bfseries 524} (2002) 153--160},
  \href{http://arxiv.org/abs/hep-ph/0110314}{{\ttfamily arXiv:hep-ph/0110314}}.

\bibitem{Stephenson:2004wv}
G.~J. Stephenson, Jr., J.~T. Goldman, B.~H.~J. McKellar, and M.~Garbutt,
  ``{Large mixing from small: PseudoDirac neutrinos and the singular seesaw},''
  \href{http://dx.doi.org/10.1142/S0217751X05028466}{{\em Int. J. Mod. Phys. A}
  {\bfseries 20} (2005) 6373--6390},
  \href{http://arxiv.org/abs/hep-ph/0404015}{{\ttfamily arXiv:hep-ph/0404015}}.

\bibitem{McDonald:2004qx}
K.~L. McDonald and B.~H.~J. McKellar, ``{The Type-II Singular See-Saw
  Mechanism},'' \href{http://dx.doi.org/10.1142/S0217751X07036567}{{\em Int. J.
  Mod. Phys. A} {\bfseries 22} (2007) 2211--2222},
  \href{http://arxiv.org/abs/hep-ph/0401073}{{\ttfamily arXiv:hep-ph/0401073}}.

\bibitem{deGouvea:2009fp}
A.~de~Gouvea, W.-C. Huang, and J.~Jenkins, ``{Pseudo-Dirac Neutrinos in the New
  Standard Model},'' \href{http://dx.doi.org/10.1103/PhysRevD.80.073007}{{\em
  Phys. Rev. D} {\bfseries 80} (2009) 073007},
  \href{http://arxiv.org/abs/0906.1611}{{\ttfamily arXiv:0906.1611 [hep-ph]}}.

\bibitem{Ahn:2016hhq}
Y.~H. Ahn, S.~K. Kang, and C.~S. Kim, ``{A Model for Pseudo-Dirac Neutrinos:
  Leptogenesis and Ultra-High Energy Neutrinos},''
  \href{http://dx.doi.org/10.1007/JHEP10(2016)092}{{\em JHEP} {\bfseries 10}
  (2016) 092}, \href{http://arxiv.org/abs/1602.05276}{{\ttfamily
  arXiv:1602.05276 [hep-ph]}}.

\bibitem{Joshipura:2013yba}
A.~S. Joshipura, S.~Mohanty, and S.~Pakvasa, ``{Pseudo-Dirac neutrinos via a
  mirror world and depletion of ultrahigh energy neutrinos},''
  \href{http://dx.doi.org/10.1103/PhysRevD.89.033003}{{\em Phys. Rev. D}
  {\bfseries 89} no.~3, (2014) 033003},
  \href{http://arxiv.org/abs/1307.5712}{{\ttfamily arXiv:1307.5712 [hep-ph]}}.

\bibitem{Babu:2022ikf}
K.~S. Babu, X.-G. He, M.~Su, and A.~Thapa, ``{Naturally light Dirac and
  pseudo-Dirac neutrinos from left-right symmetry},''
  \href{http://dx.doi.org/10.1007/JHEP08(2022)140}{{\em JHEP} {\bfseries 08}
  (2022) 140}, \href{http://arxiv.org/abs/2205.09127}{{\ttfamily
  arXiv:2205.09127 [hep-ph]}}.

\bibitem{Carloni:2022cqz}
K.~Carloni, I.~Mart\'\i{}nez-Soler, C.~A. Arguelles, K.~S. Babu, and P.~S.~B.
  Dev, ``{Probing pseudo-Dirac neutrinos with astrophysical sources at
  IceCube},'' \href{http://dx.doi.org/10.1103/PhysRevD.109.L051702}{{\em Phys.
  Rev. D} {\bfseries 109} (2024) L051702},
  \href{http://arxiv.org/abs/2212.00737}{{\ttfamily arXiv:2212.00737
  [astro-ph.HE]}}.

\bibitem{Ooguri:2016pdq}
H.~Ooguri and C.~Vafa, ``{Non-supersymmetric AdS and the Swampland},''
  \href{http://dx.doi.org/10.4310/ATMP.2017.v21.n7.a8}{{\em Adv. Theor. Math.
  Phys.} {\bfseries 21} (2017) 1787--1801},
  \href{http://arxiv.org/abs/1610.01533}{{\ttfamily arXiv:1610.01533
  [hep-th]}}.

\bibitem{Ibanez:2017kvh}
L.~E. Ibanez, V.~Martin-Lozano, and I.~Valenzuela, ``{Constraining Neutrino
  Masses, the Cosmological Constant and BSM Physics from the Weak Gravity
  Conjecture},'' \href{http://dx.doi.org/10.1007/JHEP11(2017)066}{{\em JHEP}
  {\bfseries 11} (2017) 066}, \href{http://arxiv.org/abs/1706.05392}{{\ttfamily
  arXiv:1706.05392 [hep-th]}}.

\bibitem{Gonzalo:2021zsp}
E.~Gonzalo, L.~E. Ib\'a\~nez, and I.~Valenzuela, ``{Swampland constraints on
  neutrino masses},'' \href{http://dx.doi.org/10.1007/JHEP02(2022)088}{{\em
  JHEP} {\bfseries 02} (2022) 088},
  \href{http://arxiv.org/abs/2109.10961}{{\ttfamily arXiv:2109.10961
  [hep-th]}}.

\bibitem{Casas:2024clw}
G.~F. Casas, L.~E. Ib\'a\~nez, and F.~Marchesano, ``{On small Dirac Neutrino
  Masses in String Theory},'' \href{http://arxiv.org/abs/2406.14609}{{\ttfamily
  arXiv:2406.14609 [hep-th]}}.

\bibitem{Fong:2020smz}
C.~S. Fong, T.~Gregoire, and A.~Tonero, ``{Testing quasi-Dirac leptogenesis
  through neutrino oscillations},''
  \href{http://dx.doi.org/10.1016/j.physletb.2021.136175}{{\em Phys. Lett. B}
  {\bfseries 816} (2021) 136175},
  \href{http://arxiv.org/abs/2007.09158}{{\ttfamily arXiv:2007.09158
  [hep-ph]}}.

\bibitem{Chianese:2018luo}
M.~Chianese, P.~Di~Bari, K.~Farrag, and R.~Samanta, ``{Probing relic neutrino
  radiative decays with 21 cm cosmology},''
  \href{http://dx.doi.org/10.1016/j.physletb.2018.09.040}{{\em Phys. Lett. B}
  {\bfseries 790} (2019) 64--70},
  \href{http://arxiv.org/abs/1805.11717}{{\ttfamily arXiv:1805.11717
  [hep-ph]}}.

\bibitem{Dev:2023wel}
P.~S.~B. Dev, P.~Di~Bari, I.~Mart\'\i{}nez-Soler, and R.~Roshan, ``{Relic
  neutrino decay solution to the excess radio background},''
  \href{http://dx.doi.org/10.1088/1475-7516/2024/04/046}{{\em JCAP} {\bfseries
  04} (2024) 046}, \href{http://arxiv.org/abs/2312.03082}{{\ttfamily
  arXiv:2312.03082 [hep-ph]}}.

\bibitem{Giunti:1992hk}
C.~Giunti, C.~W. Kim, and U.~W. Lee, ``{Oscillations of pseudoDirac neutrinos
  and the solar neutrino problem},''
  \href{http://dx.doi.org/10.1103/PhysRevD.46.3034}{{\em Phys. Rev. D}
  {\bfseries 46} (1992) 3034--3039},
  \href{http://arxiv.org/abs/hep-ph/9205214}{{\ttfamily arXiv:hep-ph/9205214}}.

\bibitem{Cirelli:2004cz}
M.~Cirelli, G.~Marandella, A.~Strumia, and F.~Vissani, ``{Probing oscillations
  into sterile neutrinos with cosmology, astrophysics and experiments},''
  \href{http://dx.doi.org/10.1016/j.nuclphysb.2004.11.056}{{\em Nucl. Phys. B}
  {\bfseries 708} (2005) 215--267},
  \href{http://arxiv.org/abs/hep-ph/0403158}{{\ttfamily arXiv:hep-ph/0403158}}.

\bibitem{Anamiati:2017rxw}
G.~Anamiati, R.~M. Fonseca, and M.~Hirsch, ``{Quasi Dirac neutrino
  oscillations},'' \href{http://dx.doi.org/10.1103/PhysRevD.97.095008}{{\em
  Phys. Rev. D} {\bfseries 97} no.~9, (2018) 095008},
  \href{http://arxiv.org/abs/1710.06249}{{\ttfamily arXiv:1710.06249
  [hep-ph]}}.

\bibitem{deGouvea:2021ymm}
A.~de~Gouv\^ea, E.~McGinness, I.~Martinez-Soler, and Y.~F. Perez-Gonzalez,
  ``{pp solar neutrinos at DARWIN},''
  \href{http://dx.doi.org/10.1103/PhysRevD.106.096017}{{\em Phys. Rev. D}
  {\bfseries 106} no.~9, (2022) 096017},
  \href{http://arxiv.org/abs/2111.02421}{{\ttfamily arXiv:2111.02421
  [hep-ph]}}.

\bibitem{Ansarifard:2022kvy}
S.~Ansarifard and Y.~Farzan, ``{Revisiting pseudo-Dirac neutrino scenario after
  recent solar neutrino data},''
  \href{http://dx.doi.org/10.1103/PhysRevD.107.075029}{{\em Phys. Rev. D}
  {\bfseries 107} no.~7, (2023) 075029},
  \href{http://arxiv.org/abs/2211.09105}{{\ttfamily arXiv:2211.09105
  [hep-ph]}}.

\bibitem{Franklin:2023diy}
J.~Franklin, Y.~F. Perez-Gonzalez, and J.~Turner, ``{JUNO as a Probe of the
  Pseudo-Dirac Nature using Solar Neutrinos},''
  \href{http://arxiv.org/abs/2304.05418}{{\ttfamily arXiv:2304.05418
  [hep-ph]}}.

\bibitem{DeGouvea:2020ang}
A.~De~Gouv\^ea, I.~Martinez-Soler, Y.~F. Perez-Gonzalez, and M.~Sen,
  ``{Fundamental physics with the diffuse supernova background neutrinos},''
  \href{http://dx.doi.org/10.1103/PhysRevD.102.123012}{{\em Phys. Rev. D}
  {\bfseries 102} (2020) 123012},
  \href{http://arxiv.org/abs/2007.13748}{{\ttfamily arXiv:2007.13748
  [hep-ph]}}.

\bibitem{Martinez-Soler:2021unz}
I.~Martinez-Soler, Y.~F. Perez-Gonzalez, and M.~Sen, ``{Signs of pseudo-Dirac
  neutrinos in SN1987A data},''
  \href{http://dx.doi.org/10.1103/PhysRevD.105.095019}{{\em Phys. Rev. D}
  {\bfseries 105} no.~9, (2022) 095019},
  \href{http://arxiv.org/abs/2105.12736}{{\ttfamily arXiv:2105.12736
  [hep-ph]}}.

\bibitem{2000ApJS}
R.~M. {Crocker}, F.~{Melia}, and R.~R. {Volkas}, ``{Oscillating Neutrinos from
  the Galactic Center},'' \href{http://dx.doi.org/10.1086/317350}{{\em
  Astrophys. J. Suppl. Ser.} {\bfseries 130} no.~2, (Oct., 2000) 339--350},
  \href{http://arxiv.org/abs/astro-ph/9911292}{{\ttfamily
  arXiv:astro-ph/9911292 [astro-ph]}}.

\bibitem{2002ApJS}
R.~M. {Crocker}, F.~{Melia}, and R.~R. {Volkas}, ``{Searching for
  Long-Wavelength Neutrino Oscillations in the Distorted Neutrino Spectrum of
  Galactic Supernova Remnants},'' \href{http://dx.doi.org/10.1086/340278}{{\em
  Astrophys. J. Suppl. Ser.} {\bfseries 141} no.~1, (July, 2002) 147--155},
  \href{http://arxiv.org/abs/astro-ph/0106090}{{\ttfamily
  arXiv:astro-ph/0106090 [astro-ph]}}.

\bibitem{Beacom:2003eu}
J.~F. Beacom, N.~F. Bell, D.~Hooper, J.~G. Learned, S.~Pakvasa, and T.~J.
  Weiler, ``{PseudoDirac neutrinos: A Challenge for neutrino telescopes},''
  \href{http://dx.doi.org/10.1103/PhysRevLett.92.011101}{{\em Phys. Rev. Lett.}
  {\bfseries 92} (2004) 011101},
  \href{http://arxiv.org/abs/hep-ph/0307151}{{\ttfamily arXiv:hep-ph/0307151}}.

\bibitem{Keranen:2003xd}
P.~Keranen, J.~Maalampi, M.~Myyrylainen, and J.~Riittinen, ``{Effects of
  sterile neutrinos on the ultrahigh-energy cosmic neutrino flux},''
  \href{http://dx.doi.org/10.1016/j.physletb.2003.09.006}{{\em Phys. Lett. B}
  {\bfseries 574} (2003) 162--168},
  \href{http://arxiv.org/abs/hep-ph/0307041}{{\ttfamily arXiv:hep-ph/0307041}}.

\bibitem{Esmaili:2009fk}
A.~Esmaili, ``{Pseudo-Dirac Neutrino Scenario: Cosmic Neutrinos at Neutrino
  Telescopes},'' \href{http://dx.doi.org/10.1103/PhysRevD.81.013006}{{\em Phys.
  Rev. D} {\bfseries 81} (2010) 013006},
  \href{http://arxiv.org/abs/0909.5410}{{\ttfamily arXiv:0909.5410 [hep-ph]}}.

\bibitem{Esmaili:2012ac}
A.~Esmaili and Y.~Farzan, ``{Implications of the Pseudo-Dirac Scenario for
  Ultra High Energy Neutrinos from GRBs},''
  \href{http://dx.doi.org/10.1088/1475-7516/2012/12/014}{{\em JCAP} {\bfseries
  12} (2012) 014}, \href{http://arxiv.org/abs/1208.6012}{{\ttfamily
  arXiv:1208.6012 [hep-ph]}}.

\bibitem{Shoemaker:2015qul}
I.~M. Shoemaker and K.~Murase, ``{Probing BSM Neutrino Physics with Flavor and
  Spectral Distortions: Prospects for Future High-Energy Neutrino
  Telescopes},'' \href{http://dx.doi.org/10.1103/PhysRevD.93.085004}{{\em Phys.
  Rev. D} {\bfseries 93} no.~8, (2016) 085004},
  \href{http://arxiv.org/abs/1512.07228}{{\ttfamily arXiv:1512.07228
  [astro-ph.HE]}}.

\bibitem{Brdar:2018tce}
V.~Brdar and R.~S.~L. Hansen, ``{IceCube Flavor Ratios with Identified
  Astrophysical Sources: Towards Improving New Physics Testability},''
  \href{http://dx.doi.org/10.1088/1475-7516/2019/02/023}{{\em JCAP} {\bfseries
  02} (2019) 023}, \href{http://arxiv.org/abs/1812.05541}{{\ttfamily
  arXiv:1812.05541 [hep-ph]}}.

\bibitem{Perez-Gonzalez:2023llw}
Y.~F. Perez-Gonzalez and M.~Sen, ``{From Dirac to Majorana: The cosmic neutrino
  background capture rate in the minimally extended Standard Model},''
  \href{http://dx.doi.org/10.1103/PhysRevD.109.023022}{{\em Phys. Rev. D}
  {\bfseries 109} no.~2, (2024) 023022},
  \href{http://arxiv.org/abs/2308.05147}{{\ttfamily arXiv:2308.05147
  [hep-ph]}}.

\bibitem{Barbieri:1989ti}
R.~Barbieri and A.~Dolgov, ``{Bounds on Sterile-neutrinos from
  Nucleosynthesis},''
  \href{http://dx.doi.org/10.1016/0370-2693(90)91203-N}{{\em Phys. Lett. B}
  {\bfseries 237} (1990) 440--445}.

\bibitem{Enqvist:1990ek}
K.~Enqvist, K.~Kainulainen, and J.~Maalampi, ``{Resonant neutrino transitions
  and nucleosynthesis},''
  \href{http://dx.doi.org/10.1016/0370-2693(90)91030-F}{{\em Phys. Lett. B}
  {\bfseries 249} (1990) 531--534}.

\bibitem{Chen:2022zts}
Z.~Chen, J.~Liao, J.~Ling, and B.~Yue, ``{Constraining super-light sterile
  neutrinos at Borexino and KamLAND},''
  \href{http://dx.doi.org/10.1007/JHEP09(2022)004}{{\em JHEP} {\bfseries 09}
  (2022) 004}, \href{http://arxiv.org/abs/2205.07574}{{\ttfamily
  arXiv:2205.07574 [hep-ph]}}.

\bibitem{IceCube:2022der}
{\bfseries IceCube} Collaboration, R.~Abbasi {\em et~al.}, ``{Evidence for
  neutrino emission from the nearby active galaxy NGC 1068},''
  \href{http://dx.doi.org/10.1126/science.abg3395}{{\em Science} {\bfseries
  378} no.~6619, (2022) 538--543},
  \href{http://arxiv.org/abs/2211.09972}{{\ttfamily arXiv:2211.09972
  [astro-ph.HE]}}.

\bibitem{Rink:2022nvw}
T.~Rink and M.~Sen, ``{Constraints on pseudo-Dirac neutrinos using high-energy
  neutrinos from NGC 1068},''
  \href{http://dx.doi.org/10.1016/j.physletb.2024.138558}{{\em Phys. Lett. B}
  {\bfseries 851} (2024) 138558},
  \href{http://arxiv.org/abs/2211.16520}{{\ttfamily arXiv:2211.16520
  [hep-ph]}}.

\bibitem{Dixit:2024ldv}
K.~Dixit, L.~S. Miranda, and S.~Razzaque, ``{Searching for Pseudo-Dirac
  neutrinos from Astrophysical sources in IceCube data},''
  \href{http://arxiv.org/abs/2406.06476}{{\ttfamily arXiv:2406.06476
  [astro-ph.HE]}}.

\bibitem{IceCube:2013uto}
{\bfseries IceCube} Collaboration, M.~G. Aartsen {\em et~al.}, ``{Observation
  of the cosmic-ray shadow of the Moon with IceCube},''
  \href{http://dx.doi.org/10.1103/PhysRevD.89.102004}{{\em Phys. Rev. D}
  {\bfseries 89} no.~10, (2014) 102004},
  \href{http://arxiv.org/abs/1305.6811}{{\ttfamily arXiv:1305.6811
  [astro-ph.HE]}}.

\bibitem{IceCube:2021xar}
{\bfseries IceCube} Collaboration, R.~Abbasi {\em et~al.}, ``{IceCube Data for
  Neutrino Point-Source Searches Years 2008-2018},''
  \href{http://dx.doi.org/10.21234/CPKQ-K003}{ (1, 2021) },
  \href{http://arxiv.org/abs/2101.09836}{{\ttfamily arXiv:2101.09836
  [astro-ph.HE]}}.

\bibitem{IceCube:2017der}
{\bfseries IceCube} Collaboration, M.~G. Aartsen {\em et~al.}, ``{Search for
  astrophysical sources of neutrinos using cascade events in IceCube},''
  \href{http://dx.doi.org/10.3847/1538-4357/aa8508}{{\em Astrophys. J.}
  {\bfseries 846} no.~2, (2017) 136},
  \href{http://arxiv.org/abs/1705.02383}{{\ttfamily arXiv:1705.02383
  [astro-ph.HE]}}.

\bibitem{KM3Net:2016zxf}
{\bfseries KM3Net} Collaboration, S.~Adrian-Martinez {\em et~al.}, ``{Letter of
  intent for KM3NeT 2.0}''
  \href{http://dx.doi.org/10.1088/0954-3899/43/8/084001}{{\em J. Phys. G}
  {\bfseries 43} no.~8, (2016) 084001},
  \href{http://arxiv.org/abs/1601.07459}{{\ttfamily arXiv:1601.07459
  [astro-ph.IM]}}.

\bibitem{vanEeden:2021zzv}
{\bfseries KM3NeT} Collaboration, T.~van Eeden, J.~Seneca, and A.~Heijboer,
  ``{High-energy reconstruction for single and double cascades using the KM3NeT
  detector},'' \href{http://dx.doi.org/10.22323/1.395.1089}{{\em PoS}
  {\bfseries ICRC2021} (2021) 1089},
  \href{http://arxiv.org/abs/2205.02641}{{\ttfamily arXiv:2205.02641
  [astro-ph.IM]}}.

\bibitem{Song:2020nfh}
N.~Song, S.~W. Li, C.~A. Arg\"uelles, M.~Bustamante, and A.~C. Vincent, ``{The
  Future of High-Energy Astrophysical Neutrino Flavor Measurements},''
  \href{http://dx.doi.org/10.1088/1475-7516/2021/04/054}{{\em JCAP} {\bfseries
  04} (2021) 054}, \href{http://arxiv.org/abs/2012.12893}{{\ttfamily
  arXiv:2012.12893 [hep-ph]}}.

\bibitem{Allakhverdyan:2021vkk}
V.~A. Allakhverdyan {\em et~al.}, ``{Deep-Water Neutrino Telescope in Lake
  Baikal},'' \href{http://dx.doi.org/10.1134/S1063778821090064}{{\em Phys.
  Atom. Nucl.} {\bfseries 84} no.~9, (2021) 1600--1609}.

\bibitem{IceCube-Gen2:2020qha}
{\bfseries IceCube-Gen2} Collaboration, M.~G. Aartsen {\em et~al.},
  ``{IceCube-Gen2: the window to the extreme Universe},''
  \href{http://dx.doi.org/10.1088/1361-6471/abbd48}{{\em J. Phys. G} {\bfseries
  48} no.~6, (2021) 060501}, \href{http://arxiv.org/abs/2008.04323}{{\ttfamily
  arXiv:2008.04323 [astro-ph.HE]}}.

\bibitem{P-ONE:2020ljt}
{\bfseries P-ONE} Collaboration, M.~Agostini {\em et~al.}, ``{The Pacific Ocean
  Neutrino Experiment},''
  \href{http://dx.doi.org/10.1038/s41550-020-1182-4}{{\em Nature Astron.}
  {\bfseries 4} no.~10, (2020) 913--915},
  \href{http://arxiv.org/abs/2005.09493}{{\ttfamily arXiv:2005.09493
  [astro-ph.HE]}}.

\bibitem{Zhang:2024slv}
H.~Zhang, Y.~Cui, Y.~Huang, S.~Lin, Y.~Liu, Z.~Qiu, C.~Shao, Y.~Shi, C.~Xie,
  and L.~Yang, ``{A proposed deep sea Neutrino Observatory in the Nanhai},''
  \href{http://arxiv.org/abs/2408.05122}{{\ttfamily arXiv:2408.05122
  [astro-ph.HE]}}.

\bibitem{Huang:2023mzt}
T.-Q. Huang, Z.~Cao, M.~Chen, J.~Liu, Z.~Wang, X.~You, and Y.~Qi, ``{Proposal
  for the High Energy Neutrino Telescope},''
  \href{http://dx.doi.org/10.22323/1.444.1080}{{\em PoS} {\bfseries ICRC2023}
  (2023) 1080}.

\bibitem{Ye:2022vbk}
Z.~P. Ye {\em et~al.}, ``{A multi-cubic-kilometre neutrino telescope in the
  western Pacific Ocean},'' \href{http://arxiv.org/abs/2207.04519}{{\ttfamily
  arXiv:2207.04519 [astro-ph.HE]}}.

\bibitem{Thompson:2023pnl}
{\bfseries TAMBO} Collaboration, W.~G. Thompson, ``{TAMBO: Searching for Tau
  Neutrinos in the Peruvian Andes},'' in {\em {38th International Cosmic Ray
  Conference}}.
\newblock 8, 2023.
\newblock \href{http://arxiv.org/abs/2308.09753}{{\ttfamily arXiv:2308.09753
  [astro-ph.HE]}}.

\bibitem{Otte:2019aaf}
A.~N. Otte, A.~M. Brown, M.~Doro, A.~Falcone, J.~Holder, E.~Judd, P.~Kaaret,
  M.~Mariotti, K.~Murase, and I.~Taboada, ``{Trinity: An Air-Shower Imaging
  Instrument to detect Ultrahigh Energy Neutrinos},''
  \href{http://arxiv.org/abs/1907.08727}{{\ttfamily arXiv:1907.08727
  [astro-ph.IM]}}.

\bibitem{RadarEchoTelescope:2021czz}
{\bfseries Radar Echo Telescope} Collaboration, S.~Prohira {\em et~al.},
  ``{Toward High Energy Neutrino Detection with the Radar Echo Telescope for
  Cosmic Rays (RET-CR)},'' \href{http://dx.doi.org/10.22323/1.395.1082}{{\em
  PoS} {\bfseries ICRC2021} (2021) 1082}.

\bibitem{Glashow:1960zz}
S.~L. Glashow, ``{Resonant Scattering of Antineutrinos},''
  \href{http://dx.doi.org/10.1103/PhysRev.118.316}{{\em Phys. Rev.} {\bfseries
  118} (1960) 316--317}.

\bibitem{IceCube:2021rpz}
{\bfseries IceCube} Collaboration, M.~G. Aartsen {\em et~al.}, ``{Detection of
  a particle shower at the Glashow resonance with IceCube},''
  \href{http://dx.doi.org/10.1038/s41586-021-03256-1}{{\em Nature} {\bfseries
  591} no.~7849, (2021) 220--224},
  \href{http://arxiv.org/abs/2110.15051}{{\ttfamily arXiv:2110.15051
  [hep-ex]}}. [Erratum: Nature 592, E11 (2021)].

\bibitem{Learned:1994wg}
J.~G. Learned and S.~Pakvasa, ``{Detecting tau-neutrino oscillations at PeV
  energies},'' \href{http://dx.doi.org/10.1016/0927-6505(94)00043-3}{{\em
  Astropart. Phys.} {\bfseries 3} (1995) 267--274},
\href{http://arxiv.org/abs/hep-ph/9405296}{{\ttfamily arXiv:hep-ph/9405296
  [hep-ph]}}.

\bibitem{Rachen:1998fd}
J.~P. Rachen and P.~Meszaros, ``{Photohadronic neutrinos from transients in
  astrophysical sources},''
  \href{http://dx.doi.org/10.1103/PhysRevD.58.123005}{{\em Phys. Rev. D}
  {\bfseries 58} (1998) 123005},
  \href{http://arxiv.org/abs/astro-ph/9802280}{{\ttfamily
  arXiv:astro-ph/9802280}}.

\bibitem{Kashti:2005qa}
T.~Kashti and E.~Waxman, ``{Flavoring astrophysical neutrinos: Flavor ratios
  depend on energy},''
  \href{http://dx.doi.org/10.1103/PhysRevLett.95.181101}{{\em Phys. Rev. Lett.}
  {\bfseries 95} (2005) 181101},
  \href{http://arxiv.org/abs/astro-ph/0507599}{{\ttfamily
  arXiv:astro-ph/0507599}}.

\bibitem{Kachelriess:2007tr}
M.~Kachelriess, S.~Ostapchenko, and R.~Tomas, ``{High energy neutrino yields
  from astrophysical sources. 2. Magnetized sources},''
  \href{http://dx.doi.org/10.1103/PhysRevD.77.023007}{{\em Phys. Rev. D}
  {\bfseries 77} (2008) 023007},
  \href{http://arxiv.org/abs/0708.3047}{{\ttfamily arXiv:0708.3047
  [astro-ph]}}.

\bibitem{Hummer:2010ai}
S.~Hummer, M.~Maltoni, W.~Winter, and C.~Yaguna, ``{Energy dependent neutrino
  flavor ratios from cosmic accelerators on the Hillas plot},''
  \href{http://dx.doi.org/10.1016/j.astropartphys.2010.07.003}{{\em Astropart.
  Phys.} {\bfseries 34} (2010) 205--224},
  \href{http://arxiv.org/abs/1007.0006}{{\ttfamily arXiv:1007.0006
  [astro-ph.HE]}}.

\bibitem{Winter:2014pya}
W.~Winter, ``{Describing the Observed Cosmic Neutrinos by Interactions of
  Nuclei with Matter},''
  \href{http://dx.doi.org/10.1103/PhysRevD.90.103003}{{\em Phys. Rev. D}
  {\bfseries 90} no.~10, (2014) 103003},
  \href{http://arxiv.org/abs/1407.7536}{{\ttfamily arXiv:1407.7536
  [astro-ph.HE]}}.

\bibitem{Anchordoqui:2003vc}
L.~A. Anchordoqui, H.~Goldberg, F.~Halzen, and T.~J. Weiler, ``{Galactic point
  sources of TeV antineutrinos},''
  \href{http://dx.doi.org/10.1016/j.physletb.2004.04.054}{{\em Phys. Lett. B}
  {\bfseries 593} (2004) 42},
  \href{http://arxiv.org/abs/astro-ph/0311002}{{\ttfamily
  arXiv:astro-ph/0311002}}.

\bibitem{Anchordoqui:2014pca}
L.~A. Anchordoqui, ``{Neutron $\beta$-decay as the origin of
  IceCube\textquoteright{}s PeV (anti)neutrinos},''
  \href{http://dx.doi.org/10.1103/PhysRevD.91.027301}{{\em Phys. Rev. D}
  {\bfseries 91} (2015) 027301},
  \href{http://arxiv.org/abs/1411.6457}{{\ttfamily arXiv:1411.6457
  [astro-ph.HE]}}.

\bibitem{Lipari:2007su}
P.~Lipari, M.~Lusignoli, and D.~Meloni, ``{Flavor Composition and Energy
  Spectrum of Astrophysical Neutrinos},''
  \href{http://dx.doi.org/10.1103/PhysRevD.75.123005}{{\em Phys. Rev. D}
  {\bfseries 75} (2007) 123005},
  \href{http://arxiv.org/abs/0704.0718}{{\ttfamily arXiv:0704.0718
  [astro-ph]}}.

\bibitem{Mena:2014sja}
O.~Mena, S.~Palomares-Ruiz, and A.~C. Vincent, ``{Flavor Composition of the
  High-Energy Neutrino Events in IceCube},''
  \href{http://dx.doi.org/10.1103/PhysRevLett.113.091103}{{\em Phys. Rev.
  Lett.} {\bfseries 113} (2014) 091103},
  \href{http://arxiv.org/abs/1404.0017}{{\ttfamily arXiv:1404.0017
  [astro-ph.HE]}}.

\bibitem{Palladino:2015zua}
A.~Palladino, G.~Pagliaroli, F.~L. Villante, and F.~Vissani, ``{What is the
  Flavor of the Cosmic Neutrinos Seen by IceCube?},''
  \href{http://dx.doi.org/10.1103/PhysRevLett.114.171101}{{\em Phys. Rev.
  Lett.} {\bfseries 114} no.~17, (2015) 171101},
  \href{http://arxiv.org/abs/1502.02923}{{\ttfamily arXiv:1502.02923
  [astro-ph.HE]}}.

\bibitem{Bustamante:2015waa}
M.~Bustamante, J.~F. Beacom, and W.~Winter, ``{Theoretically palatable flavor
  combinations of astrophysical neutrinos},''
  \href{http://dx.doi.org/10.1103/PhysRevLett.115.161302}{{\em Phys. Rev.
  Lett.} {\bfseries 115} no.~16, (2015) 161302},
  \href{http://arxiv.org/abs/1506.02645}{{\ttfamily arXiv:1506.02645
  [astro-ph.HE]}}.

\bibitem{Bustamante:2019sdb}
M.~Bustamante and M.~Ahlers, ``{Inferring the flavor of high-energy
  astrophysical neutrinos at their sources},''
  \href{http://dx.doi.org/10.1103/PhysRevLett.122.241101}{{\em Phys. Rev.
  Lett.} {\bfseries 122} no.~24, (2019) 241101},
  \href{http://arxiv.org/abs/1901.10087}{{\ttfamily arXiv:1901.10087
  [astro-ph.HE]}}.

\bibitem{Palladino:2019pid}
A.~Palladino, ``{The flavor composition of astrophysical neutrinos after 8
  years of IceCube: an indication of neutron decay scenario?},''
  \href{http://dx.doi.org/10.1140/epjc/s10052-019-7018-7}{{\em Eur. Phys. J. C}
  {\bfseries 79} no.~6, (2019) 500},
  \href{http://arxiv.org/abs/1902.08630}{{\ttfamily arXiv:1902.08630
  [astro-ph.HE]}}.

\bibitem{Liu:2023flr}
Q.~Liu, D.~F.~G. Fiorillo, C.~A. Arg\"uelles, M.~Bustamante, N.~Song, and A.~C.
  Vincent, ``{Identifying Energy-Dependent Flavor Transitions in High-Energy
  Astrophysical Neutrino Measurements},''
  \href{http://arxiv.org/abs/2312.07649}{{\ttfamily arXiv:2312.07649
  [astro-ph.HE]}}.

\bibitem{Dev:2023znd}
P.~S.~B. Dev, S.~Jana, and Y.~Porto, ``{Flavor Matters, but Matter Flavors:
  Matter Effects on Flavor Composition of Astrophysical Neutrinos},''
  \href{http://arxiv.org/abs/2312.17315}{{\ttfamily arXiv:2312.17315
  [hep-ph]}}.

\bibitem{Wolfenstein:1977ue}
L.~Wolfenstein, ``{Neutrino Oscillations in Matter},''
  \href{http://dx.doi.org/10.1103/PhysRevD.17.2369}{{\em Phys. Rev. D}
  {\bfseries 17} (1978) 2369--2374}.

\bibitem{Mikheyev:1985zog}
S.~P. Mikheyev and A.~Y. Smirnov, ``{Resonance Amplification of Oscillations in
  Matter and Spectroscopy of Solar Neutrinos},'' {\em Sov. J. Nucl. Phys.}
  {\bfseries 42} (1985) 913--917.

\bibitem{ParticleDataGroup:2022pth}
{\bfseries Particle Data Group} Collaboration, R.~L. Workman {\em et~al.},
  ``{Review of Particle Physics},''
  \href{http://dx.doi.org/10.1093/ptep/ptac097}{{\em PTEP} {\bfseries 2022}
  (2022) 083C01}.

\bibitem{Esteban:2020cvm}
I.~Esteban, M.~C. Gonzalez-Garcia, M.~Maltoni, T.~Schwetz, and A.~Zhou, ``{The
  fate of hints: updated global analysis of three-flavor neutrino
  oscillations},'' \href{http://dx.doi.org/10.1007/JHEP09(2020)178}{{\em JHEP}
  {\bfseries 09} (2020) 178}, \href{http://arxiv.org/abs/2007.14792}{{\ttfamily
  arXiv:2007.14792 [hep-ph]}}.

\bibitem{nufit}
{\bfseries NuFIT} Collaboration, ``Three-neutrino fit based on data available
  in march 2024.''
\newblock \url{www.nu-fit.org}.

\bibitem{t2k}
C.~Giganti, ``T2k recent results and plans.''
\newblock
  \url{https://agenda.infn.it/event/37867/contributions/233954/attachments/121809/177671/Neutrino2024_T2K_Claudio.pdf}.
  Talk given at Neutrino 2024 (June 17, 2024).

\bibitem{nova}
J.~Wolcott, ``New nova results with 10 years of data.''
\newblock
  \url{https://agenda.infn.it/event/37867/contributions/233955/attachments/121832/177712/2024-06-17%20Wolcott%20NOvA%202024%20results%20-%20NEUTRINO.pdf}.
  Talk given at Neutrino 2024 (June 17, 2024).

\bibitem{Kelly:2020fkv}
K.~J. Kelly, P.~A.~N. Machado, S.~J. Parke, Y.~F. Perez-Gonzalez, and R.~Z.
  Funchal, ``{Neutrino mass ordering in light of recent data},''
  \href{http://dx.doi.org/10.1103/PhysRevD.103.013004}{{\em Phys. Rev. D}
  {\bfseries 103} no.~1, (2021) 013004},
  \href{http://arxiv.org/abs/2007.08526}{{\ttfamily arXiv:2007.08526
  [hep-ph]}}.

\bibitem{Anamiati:2019maf}
G.~Anamiati, V.~De~Romeri, M.~Hirsch, C.~A. Ternes, and M.~T\'ortola,
  ``{Quasi-Dirac neutrino oscillations at DUNE and JUNO},''
  \href{http://dx.doi.org/10.1103/PhysRevD.100.035032}{{\em Phys. Rev. D}
  {\bfseries 100} no.~3, (2019) 035032},
  \href{http://arxiv.org/abs/1907.00980}{{\ttfamily arXiv:1907.00980
  [hep-ph]}}.

\bibitem{Valera:2024buc}
V.~B. Valera, D.~F.~G. Fiorillo, I.~Esteban, and M.~Bustamante, ``{New limits
  on neutrino decay from high-energy astrophysical neutrinos},''
  \href{http://arxiv.org/abs/2405.14826}{{\ttfamily arXiv:2405.14826
  [astro-ph.HE]}}.

\bibitem{Batell:2024hzo}
B.~Batell and W.~Yin, ``{Cosmic Stability of Dark Matter from Pauli
  Blocking},'' \href{http://arxiv.org/abs/2406.17028}{{\ttfamily
  arXiv:2406.17028 [hep-ph]}}.

\bibitem{Berryman:2022hds}
J.~M. Berryman {\em et~al.}, ``{Neutrino self-interactions: A white paper},''
  \href{http://dx.doi.org/10.1016/j.dark.2023.101267}{{\em Phys. Dark Univ.}
  {\bfseries 42} (2023) 101267},
  \href{http://arxiv.org/abs/2203.01955}{{\ttfamily arXiv:2203.01955
  [hep-ph]}}.

\bibitem{Notzold:1987ik}
D.~N\"otzold and G.~Raffelt, ``{Neutrino dispersion at finite temperature and
  density},'' \href{http://dx.doi.org/10.1016/0550-3213(88)90113-7}{{\em Nucl.
  Phys. B} {\bfseries 307} (1988) 924--936}.

\bibitem{Stodolsky:1974aq}
L.~Stodolsky, ``{Speculations on Detection of the Neutrino Sea},''
  \href{http://dx.doi.org/10.1103/PhysRevLett.34.110}{{\em Phys. Rev. Lett.}
  {\bfseries 34} (1975) 110}. [Erratum: Phys.Rev.Lett. 34, 508 (1975)].

\bibitem{Serpico:2005bc}
P.~D. Serpico and G.~G. Raffelt, ``{Lepton asymmetry and primordial
  nucleosynthesis in the era of precision cosmology},''
  \href{http://dx.doi.org/10.1103/PhysRevD.71.127301}{{\em Phys. Rev. D}
  {\bfseries 71} (2005) 127301},
  \href{http://arxiv.org/abs/astro-ph/0506162}{{\ttfamily
  arXiv:astro-ph/0506162}}.

\bibitem{Matsumoto:2022tlr}
A.~Matsumoto {\em et~al.}, ``{EMPRESS. VIII. A New Determination of Primordial
  He Abundance with Extremely Metal-poor Galaxies: A Suggestion of the Lepton
  Asymmetry and Implications for the Hubble Tension},''
  \href{http://dx.doi.org/10.3847/1538-4357/ac9ea1}{{\em Astrophys. J.}
  {\bfseries 941} no.~2, (2022) 167},
  \href{http://arxiv.org/abs/2203.09617}{{\ttfamily arXiv:2203.09617
  [astro-ph.CO]}}.

\bibitem{Burns:2022hkq}
A.-K. Burns, T.~M.~P. Tait, and M.~Valli, ``{Indications for a Nonzero Lepton
  Asymmetry from Extremely Metal-Poor Galaxies},''
  \href{http://dx.doi.org/10.1103/PhysRevLett.130.131001}{{\em Phys. Rev.
  Lett.} {\bfseries 130} no.~13, (2023) 131001},
  \href{http://arxiv.org/abs/2206.00693}{{\ttfamily arXiv:2206.00693
  [hep-ph]}}.

\bibitem{Escudero:2022okz}
M.~Escudero, A.~Ibarra, and V.~Maura, ``{Primordial lepton asymmetries in the
  precision cosmology era: Current status and future sensitivities from BBN and
  the CMB},'' \href{http://dx.doi.org/10.1103/PhysRevD.107.035024}{{\em Phys.
  Rev. D} {\bfseries 107} no.~3, (2023) 035024},
  \href{http://arxiv.org/abs/2208.03201}{{\ttfamily arXiv:2208.03201
  [hep-ph]}}.

\bibitem{Li:2024gzf}
Y.-Z. Li and J.-H. Yu, ``{Revisiting primordial neutrino asymmetries, spectral
  distortions and cosmological constraints with full neutrino transport},''
  \href{http://arxiv.org/abs/2409.08280}{{\ttfamily arXiv:2409.08280
  [hep-ph]}}.

\bibitem{Simha:2008mt}
V.~Simha and G.~Steigman, ``{Constraining The Universal Lepton Asymmetry},''
  \href{http://dx.doi.org/10.1088/1475-7516/2008/08/011}{{\em JCAP} {\bfseries
  08} (2008) 011}, \href{http://arxiv.org/abs/0806.0179}{{\ttfamily
  arXiv:0806.0179 [hep-ph]}}.

\bibitem{Froustey:2021azz}
J.~Froustey and C.~Pitrou, ``{Primordial neutrino asymmetry evolution with full
  mean-field effects and collisions},''
  \href{http://dx.doi.org/10.1088/1475-7516/2022/03/065}{{\em JCAP} {\bfseries
  03} no.~03, (2022) 065}, \href{http://arxiv.org/abs/2110.11889}{{\ttfamily
  arXiv:2110.11889 [hep-ph]}}.

\bibitem{Froustey:2024mgf}
J.~Froustey and C.~Pitrou, ``{Constraints on primordial lepton asymmetries with
  full neutrino transport},'' \href{http://arxiv.org/abs/2405.06509}{{\ttfamily
  arXiv:2405.06509 [hep-ph]}}.

\bibitem{Kuzmin:1985mm}
V.~A. Kuzmin, V.~A. Rubakov, and M.~E. Shaposhnikov, ``{On the Anomalous
  Electroweak Baryon Number Nonconservation in the Early Universe},''
  \href{http://dx.doi.org/10.1016/0370-2693(85)91028-7}{{\em Phys. Lett. B}
  {\bfseries 155} (1985) 36}.

\bibitem{Harvey:1990qw}
J.~A. Harvey and M.~S. Turner, ``{Cosmological baryon and lepton number in the
  presence of electroweak fermion number violation},''
  \href{http://dx.doi.org/10.1103/PhysRevD.42.3344}{{\em Phys. Rev. D}
  {\bfseries 42} (1990) 3344--3349}.

\bibitem{Planck:2018vyg}
{\bfseries Planck} Collaboration, N.~Aghanim {\em et~al.}, ``{Planck 2018
  results. VI. Cosmological parameters},''
  \href{http://dx.doi.org/10.1051/0004-6361/201833910}{{\em Astron. Astrophys.}
  {\bfseries 641} (2020) A6}, \href{http://arxiv.org/abs/1807.06209}{{\ttfamily
  arXiv:1807.06209 [astro-ph.CO]}}. [Erratum: Astron.Astrophys. 652, C4
  (2021)].

\bibitem{Affleck:1984fy}
I.~Affleck and M.~Dine, ``{A New Mechanism for Baryogenesis},''
  \href{http://dx.doi.org/10.1016/0550-3213(85)90021-5}{{\em Nucl. Phys. B}
  {\bfseries 249} (1985) 361--380}.

\bibitem{Dolgov:1989us}
A.~D. Dolgov and D.~P. Kirilova, ``{ON PARTICLE CREATION BY A TIME DEPENDENT
  SCALAR FIELD},'' {\em Sov. J. Nucl. Phys.} {\bfseries 51} (1990) 172--177.

\bibitem{Foot:1995qk}
R.~Foot, M.~J. Thomson, and R.~R. Volkas, ``{Large neutrino asymmetries from
  neutrino oscillations},''
  \href{http://dx.doi.org/10.1103/PhysRevD.53.R5349}{{\em Phys. Rev. D}
  {\bfseries 53} (1996) R5349--R5353},
  \href{http://arxiv.org/abs/hep-ph/9509327}{{\ttfamily arXiv:hep-ph/9509327}}.

\bibitem{Casas:1997gx}
A.~Casas, W.~Y. Cheng, and G.~Gelmini, ``{Generation of large lepton
  asymmetries},'' \href{http://dx.doi.org/10.1016/S0550-3213(98)00606-3}{{\em
  Nucl. Phys. B} {\bfseries 538} (1999) 297--308},
  \href{http://arxiv.org/abs/hep-ph/9709289}{{\ttfamily arXiv:hep-ph/9709289}}.

\bibitem{Bajc:1997ky}
B.~Bajc, A.~Riotto, and G.~Senjanovic, ``{Large lepton number of the universe
  and the fate of topological defects},''
  \href{http://dx.doi.org/10.1103/PhysRevLett.81.1355}{{\em Phys. Rev. Lett.}
  {\bfseries 81} (1998) 1355--1358},
  \href{http://arxiv.org/abs/hep-ph/9710415}{{\ttfamily arXiv:hep-ph/9710415}}.

\bibitem{Akhmedov:1998qx}
E.~K. Akhmedov, V.~A. Rubakov, and A.~Y. Smirnov, ``{Baryogenesis via neutrino
  oscillations},'' \href{http://dx.doi.org/10.1103/PhysRevLett.81.1359}{{\em
  Phys. Rev. Lett.} {\bfseries 81} (1998) 1359--1362},
  \href{http://arxiv.org/abs/hep-ph/9803255}{{\ttfamily arXiv:hep-ph/9803255}}.

\bibitem{McDonald:1999in}
J.~McDonald, ``{Naturally large cosmological neutrino asymmetries in the
  MSSM},'' \href{http://dx.doi.org/10.1103/PhysRevLett.84.4798}{{\em Phys. Rev.
  Lett.} {\bfseries 84} (2000) 4798--4801},
  \href{http://arxiv.org/abs/hep-ph/9908300}{{\ttfamily arXiv:hep-ph/9908300}}.

\bibitem{March-Russell:1999hpw}
J.~March-Russell, H.~Murayama, and A.~Riotto, ``{The Small observed baryon
  asymmetry from a large lepton asymmetry},''
  \href{http://dx.doi.org/10.1088/1126-6708/1999/11/015}{{\em JHEP} {\bfseries
  11} (1999) 015}, \href{http://arxiv.org/abs/hep-ph/9908396}{{\ttfamily
  arXiv:hep-ph/9908396}}.

\bibitem{Kawasaki:2022hvx}
M.~Kawasaki and K.~Murai, ``{Lepton asymmetric universe},''
  \href{http://dx.doi.org/10.1088/1475-7516/2022/08/041}{{\em JCAP} {\bfseries
  08} no.~08, (2022) 041}, \href{http://arxiv.org/abs/2203.09713}{{\ttfamily
  arXiv:2203.09713 [hep-ph]}}.

\bibitem{Domcke:2022uue}
V.~Domcke, K.~Kamada, K.~Mukaida, K.~Schmitz, and M.~Yamada, ``{New Constraint
  on Primordial Lepton Flavor Asymmetries},''
  \href{http://dx.doi.org/10.1103/PhysRevLett.130.261803}{{\em Phys. Rev.
  Lett.} {\bfseries 130} no.~26, (2023) 261803},
  \href{http://arxiv.org/abs/2208.03237}{{\ttfamily arXiv:2208.03237
  [hep-ph]}}.

\bibitem{Borah:2022uos}
D.~Borah and A.~Dasgupta, ``{Large neutrino asymmetry from TeV scale
  leptogenesis},'' \href{http://dx.doi.org/10.1103/PhysRevD.108.035015}{{\em
  Phys. Rev. D} {\bfseries 108} no.~3, (2023) 035015},
  \href{http://arxiv.org/abs/2206.14722}{{\ttfamily arXiv:2206.14722
  [hep-ph]}}.

\bibitem{ChoeJo:2023cnx}
Y.~ChoeJo, K.~Enomoto, Y.~Kim, and H.-S. Lee, ``{Second leptogenesis:
  Unraveling the baryon-lepton asymmetry discrepancy},''
  \href{http://dx.doi.org/10.1007/JHEP03(2024)003}{{\em JHEP} {\bfseries 03}
  (2024) 003}, \href{http://arxiv.org/abs/2311.16672}{{\ttfamily
  arXiv:2311.16672 [hep-ph]}}.

\bibitem{Borah:2024xoa}
D.~Borah, N.~Das, and I.~Saha, ``{Large neutrino asymmetry from forbidden decay
  of dark matter},'' \href{http://arxiv.org/abs/2410.00096}{{\ttfamily
  arXiv:2410.00096 [hep-ph]}}.

\bibitem{Shi:1998km}
X.-D. Shi and G.~M. Fuller, ``{A New dark matter candidate: Nonthermal sterile
  neutrinos},'' \href{http://dx.doi.org/10.1103/PhysRevLett.82.2832}{{\em Phys.
  Rev. Lett.} {\bfseries 82} (1999) 2832--2835},
  \href{http://arxiv.org/abs/astro-ph/9810076}{{\ttfamily
  arXiv:astro-ph/9810076}}.

\bibitem{Dalton:2021afc}
T.~Dalton, S.~L. Morris, and M.~Fumagalli, ``{Probing the physical properties
  of the intergalactic medium using gamma-ray bursts},''
  \href{http://dx.doi.org/10.1093/mnras/stab335}{{\em Mon. Not. Roy. Astron.
  Soc.} {\bfseries 502} no.~4, (2021) 5981--5996},
  \href{http://arxiv.org/abs/2102.02530}{{\ttfamily arXiv:2102.02530
  [astro-ph.CO]}}.

\bibitem{Hopkins:2006bw}
A.~M. Hopkins and J.~F. Beacom, ``{On the normalisation of the cosmic star
  formation history},'' \href{http://dx.doi.org/10.1086/506610}{{\em Astrophys.
  J.} {\bfseries 651} (2006) 142--154},
  \href{http://arxiv.org/abs/astro-ph/0601463}{{\ttfamily
  arXiv:astro-ph/0601463}}.

\bibitem{Yuksel:2008cu}
H.~Yuksel, M.~D. Kistler, J.~F. Beacom, and A.~M. Hopkins, ``{Revealing the
  High-Redshift Star Formation Rate with Gamma-Ray Bursts},''
  \href{http://dx.doi.org/10.1086/591449}{{\em Astrophys. J. Lett.} {\bfseries
  683} (2008) L5--L8}, \href{http://arxiv.org/abs/0804.4008}{{\ttfamily
  arXiv:0804.4008 [astro-ph]}}.

\bibitem{IceCube:2023ame}
{\bfseries IceCube} Collaboration, R.~Abbasi {\em et~al.}, ``{Observation of
  high-energy neutrinos from the Galactic plane},''
  \href{http://dx.doi.org/10.1126/science.adc9818}{{\em Science} {\bfseries
  380} no.~6652, (2023) adc9818},
  \href{http://arxiv.org/abs/2307.04427}{{\ttfamily arXiv:2307.04427
  [astro-ph.HE]}}.

\bibitem{KATRIN:2022kkv}
{\bfseries KATRIN} Collaboration, M.~Aker {\em et~al.}, ``{New Constraint on
  the Local Relic Neutrino Background Overdensity with the First KATRIN Data
  Runs},'' \href{http://dx.doi.org/10.1103/PhysRevLett.129.011806}{{\em Phys.
  Rev. Lett.} {\bfseries 129} no.~1, (2022) 011806},
  \href{http://arxiv.org/abs/2202.04587}{{\ttfamily arXiv:2202.04587
  [nucl-ex]}}.

\bibitem{Brdar:2022kpu}
V.~Brdar, P.~S.~B. Dev, R.~Plestid, and A.~Soni, ``{A new probe of relic
  neutrino clustering using cosmogenic neutrinos},''
  \href{http://dx.doi.org/10.1016/j.physletb.2022.137358}{{\em Phys. Lett. B}
  {\bfseries 833} (2022) 137358},
  \href{http://arxiv.org/abs/2207.02860}{{\ttfamily arXiv:2207.02860
  [hep-ph]}}.

\bibitem{Bauer:2022lri}
M.~Bauer and J.~D. Shergold, ``{Limits on the cosmic neutrino background},''
  \href{http://dx.doi.org/10.1088/1475-7516/2023/01/003}{{\em JCAP} {\bfseries
  01} (2023) 003}, \href{http://arxiv.org/abs/2207.12413}{{\ttfamily
  arXiv:2207.12413 [hep-ph]}}.

\bibitem{Tsai:2022jnv}
Y.-D. Tsai, J.~Eby, J.~Arakawa, D.~Farnocchia, and M.~S. Safronova,
  ``{OSIRIS-REx constraints on local dark matter and cosmic neutrino
  profiles},'' \href{http://dx.doi.org/10.1088/1475-7516/2024/02/029}{{\em
  JCAP} {\bfseries 02} (2024) 029},
  \href{http://arxiv.org/abs/2210.03749}{{\ttfamily arXiv:2210.03749
  [hep-ph]}}.

\bibitem{Ciscar-Monsalvatje:2024tvm}
M.~C\'\i{}scar-Monsalvatje, G.~Herrera, and I.~M. Shoemaker, ``{Upper Limits on
  the Cosmic Neutrino Background from Cosmic Rays},''
  \href{http://arxiv.org/abs/2402.00985}{{\ttfamily arXiv:2402.00985
  [hep-ph]}}.

\bibitem{Franklin:2024enc}
J.~Franklin, I.~Martinez-Soler, Y.~F. Perez-Gonzalez, and J.~Turner,
  ``{Constraints on the Cosmic Neutrino Background from NGC 1068},''
  \href{http://arxiv.org/abs/2404.02202}{{\ttfamily arXiv:2404.02202
  [hep-ph]}}.

\bibitem{DeMarchi:2024zer}
A.~G. De~Marchi, A.~Granelli, J.~Nava, and F.~Sala, ``{Relic Neutrino
  Background from Cosmic-Ray Reservoirs},''
  \href{http://arxiv.org/abs/2405.04568}{{\ttfamily arXiv:2405.04568
  [hep-ph]}}.

\bibitem{Herrera:2024upj}
G.~Herrera, S.~Horiuchi, and X.~Qi, ``{Diffuse Boosted Cosmic Neutrino
  Background},'' \href{http://arxiv.org/abs/2405.14946}{{\ttfamily
  arXiv:2405.14946 [hep-ph]}}.

\bibitem{Ringwald:2004np}
A.~Ringwald and Y.~Y.~Y. Wong, ``{Gravitational clustering of relic neutrinos
  and implications for their detection},''
  \href{http://dx.doi.org/10.1088/1475-7516/2004/12/005}{{\em JCAP} {\bfseries
  12} (2004) 005}, \href{http://arxiv.org/abs/hep-ph/0408241}{{\ttfamily
  arXiv:hep-ph/0408241}}.

\bibitem{Mertsch:2019qjv}
P.~Mertsch, G.~Parimbelli, P.~F. de~Salas, S.~Gariazzo, J.~Lesgourgues, and
  S.~Pastor, ``{Neutrino clustering in the Milky Way and beyond},''
  \href{http://dx.doi.org/10.1088/1475-7516/2020/01/015}{{\em JCAP} {\bfseries
  01} (2020) 015}, \href{http://arxiv.org/abs/1910.13388}{{\ttfamily
  arXiv:1910.13388 [astro-ph.CO]}}.

\bibitem{Zimmer:2023jbb}
F.~Zimmer, C.~A. Correa, and S.~Ando, ``{Influence of local structure on relic
  neutrino abundances and anisotropies},''
  \href{http://dx.doi.org/10.1088/1475-7516/2023/11/038}{{\em JCAP} {\bfseries
  11} (2023) 038}, \href{http://arxiv.org/abs/2306.16444}{{\ttfamily
  arXiv:2306.16444 [astro-ph.CO]}}.

\bibitem{Elbers:2023mdr}
W.~Elbers, C.~S. Frenk, A.~Jenkins, B.~Li, S.~Pascoli, J.~Jasche, G.~Lavaux,
  and V.~Springel, ``{Where shadows lie: reconstruction of anisotropies in the
  neutrino sky},'' \href{http://dx.doi.org/10.1088/1475-7516/2023/10/010}{{\em
  JCAP} {\bfseries 10} (2023) 010},
  \href{http://arxiv.org/abs/2307.03191}{{\ttfamily arXiv:2307.03191
  [astro-ph.CO]}}.

\bibitem{Holm:2024zpr}
E.~B. Holm, S.~Zentarra, and I.~M. Oldengott, ``{Local clustering of relic
  neutrinos: Comparison of kinetic field theory and the Vlasov equation},''
  \href{http://arxiv.org/abs/2404.11295}{{\ttfamily arXiv:2404.11295
  [hep-ph]}}.

\bibitem{Bondarenko:2023ukx}
K.~Bondarenko, A.~Boyarsky, J.~Pradler, and A.~Sokolenko, ``{Best-case
  scenarios for neutrino capture experiments},''
  \href{http://dx.doi.org/10.1088/1475-7516/2023/10/026}{{\em JCAP} {\bfseries
  10} (2023) 026}, \href{http://arxiv.org/abs/2306.12366}{{\ttfamily
  arXiv:2306.12366 [hep-ph]}}.

\bibitem{Smirnov:2022sfo}
A.~Y. Smirnov and X.-J. Xu, ``{Neutrino bound states and bound systems},''
  \href{http://dx.doi.org/10.1007/JHEP08(2022)170}{{\em JHEP} {\bfseries 08}
  (2022) 170}, \href{http://arxiv.org/abs/2201.00939}{{\ttfamily
  arXiv:2201.00939 [hep-ph]}}.

\bibitem{1969Ap......5...67E}
Y.~I. {Einasto}, ``{The andromeda galaxy M 31: I. A preliminary model},''
  \href{http://dx.doi.org/10.1007/BF01013353}{{\em Astrophysics} {\bfseries 5}
  (1969) 67--80}.

\bibitem{Fang:2022trf}
K.~Fang, J.~S. Gallagher, and F.~Halzen, ``{The TeV Diffuse Cosmic Neutrino
  Spectrum and the Nature of Astrophysical Neutrino Sources},''
  \href{http://dx.doi.org/10.3847/1538-4357/ac7649}{{\em Astrophys. J.}
  {\bfseries 933} no.~2, (2022) 190},
  \href{http://arxiv.org/abs/2205.03740}{{\ttfamily arXiv:2205.03740
  [astro-ph.HE]}}.

\bibitem{IceCube:2024nhk}
{\bfseries IceCube} Collaboration, R.~Abbasi {\em et~al.}, ``{Observation of
  Seven Astrophysical Tau Neutrino Candidates with IceCube},''
  \href{http://dx.doi.org/10.1103/PhysRevLett.132.151001}{{\em Phys. Rev.
  Lett.} {\bfseries 132} no.~15, (2024) 151001},
  \href{http://arxiv.org/abs/2403.02516}{{\ttfamily arXiv:2403.02516
  [astro-ph.HE]}}.

\bibitem{IceCube:2015gsk}
{\bfseries IceCube} Collaboration, M.~G. Aartsen {\em et~al.}, ``{A combined
  maximum-likelihood analysis of the high-energy astrophysical neutrino flux
  measured with IceCube},''
  \href{http://dx.doi.org/10.1088/0004-637X/809/1/98}{{\em Astrophys. J.}
  {\bfseries 809} no.~1, (2015) 98},
  \href{http://arxiv.org/abs/1507.03991}{{\ttfamily arXiv:1507.03991
  [astro-ph.HE]}}.

\bibitem{IceCube:2023fgt}
{\bfseries IceCube} Collaboration, D.~Cowen {\em et~al.}, ``{Summary of IceCube
  tau neutrino searches and flavor composition measurements of the diffuse
  astrophysical neutrino flux},''
  \href{http://dx.doi.org/10.22323/1.444.1122}{{\em PoS} {\bfseries ICRC2023}
  (2023) 1122}.

\bibitem{Ackermann:2022rqc}
M.~Ackermann {\em et~al.}, ``{High-energy and ultra-high-energy neutrinos: A
  Snowmass white paper},''
  \href{http://dx.doi.org/10.1016/j.jheap.2022.08.001}{{\em JHEAp} {\bfseries
  36} (2022) 55--110}, \href{http://arxiv.org/abs/2203.08096}{{\ttfamily
  arXiv:2203.08096 [hep-ph]}}.

\bibitem{IceCube-Gen2:2023rds}
{\bfseries IceCube-Gen2} Collaboration, R.~Abbasi {\em et~al.}, ``{Sensitivity
  of IceCube-Gen2 to measure flavor composition of Astrophysical neutrinos},''
  \href{http://dx.doi.org/10.22323/1.444.1123}{{\em PoS} {\bfseries ICRC2023}
  (2023) 1123}.

\bibitem{JUNO:2015zny}
{\bfseries JUNO} Collaboration, F.~An {\em et~al.}, ``{Neutrino Physics with
  JUNO},'' \href{http://dx.doi.org/10.1088/0954-3899/43/3/030401}{{\em J. Phys.
  G} {\bfseries 43} no.~3, (2016) 030401},
  \href{http://arxiv.org/abs/1507.05613}{{\ttfamily arXiv:1507.05613
  [physics.ins-det]}}.

\bibitem{DUNE:2015lol}
{\bfseries DUNE} Collaboration, R.~Acciarri {\em et~al.}, ``{Long-Baseline
  Neutrino Facility (LBNF) and Deep Underground Neutrino Experiment (DUNE)}:
  {Conceptual Design Report, Volume 2: The Physics Program for DUNE at LBNF},''
  \href{http://arxiv.org/abs/1512.06148}{{\ttfamily arXiv:1512.06148
  [physics.ins-det]}}.

\bibitem{Hyper-Kamiokande:2018ofw}
{\bfseries Hyper-Kamiokande} Collaboration, K.~Abe {\em et~al.},
  ``{Hyper-Kamiokande Design Report},''
  \href{http://arxiv.org/abs/1805.04163}{{\ttfamily arXiv:1805.04163
  [physics.ins-det]}}.

\bibitem{Fong:2024mqz}
C.~S. Fong and Y.~Porto, ``{Constraining pseudo-Diracness with astrophysical
  neutrino flavors},'' \href{http://arxiv.org/abs/2406.15566}{{\ttfamily
  arXiv:2406.15566 [hep-ph]}}.

\end{thebibliography}\endgroup

\end{document}